\documentclass[reprint,prl,superscriptaddress]{revtex4-2}

\usepackage{graphicx}
\usepackage{siunitx}
\usepackage{amsmath}
\usepackage{amsfonts}
\usepackage{amssymb}
\usepackage[normalem]{ulem}
\usepackage{hyperref}

\hypersetup{
    colorlinks=true,
    linkcolor=blue,
    citecolor=blue,
    urlcolor=blue,
}

\usepackage{xcolor}
\begin{document}

\title{Soliton-Lattice Origin of Incommensurability in High-Pressure Antimony}

\author{M. Smirman}
\affiliation{Department of Physics and Astronomy, George Mason University, Fairfax, VA 22030, USA}
\author{K. Sun}
\affiliation{Department of Physics, University of Michigan, Ann Arbor, Michigan 48109, USA}
\author{H. Sheng}
\affiliation{Department of Physics and Astronomy, George Mason University, Fairfax, VA 22030, USA}
\author{I. I. Mazin}
\affiliation{Department of Physics and Astronomy, George Mason University, Fairfax, VA 22030, USA}
\affiliation{Quantum Science and Engineering Center, George Mason University, Fairfax, VA 22030, USA}

\begin{abstract}
The microscopic origin of incommensurability in host--guest crystals remains unresolved. Here we show that the high-pressure antimony host--guest phase Sb-II is a soliton lattice, where localized guest--host misregistries decorate a commensurate 4/3 reference structure. First-principles calculations reveal a corrugated and hysteretic registry landscape, motivating a Frenkel--Kontorova description that accounts for soliton-lattice approximants of the form $(4n+1)/(3n+1)$. Machine-learning-potential calculations of these approximants show an energetically preferred soliton spacing, and their localized misregistry cores are also verified by first-principles relaxations.At finite temperature, the thermally broadened soliton lattice appears as an atomic density wave along the guest chains, consistent with experimental diffraction. This soliton-registry picture provides a microscopic mechanism for host--guest incommensurability in Sb-II.
\end{abstract}

\maketitle

Incommensurate crystals are a distinctive form of ordered matter in which long-range structural order exists without being reducible to a single three-dimensional lattice periodicity.\citep{janssen2006incommensurate,vansmaalen2007incommensurate} Antimony, like several other elements including bismuth (Bi-III) and scandium (Sc-II), exhibits an unusual high-pressure host--guest phase, Sb-II, commonly described as a host--guest phase or incommensurate composite.\citep{Schwarz_2003,McMahon_2007,McMahonNelmes2004Review,Fujihisa2005} In such structures, the crystal can be partitioned into two interpenetrating subsystems: a host framework and a guest substructure. Each subsystem is individually periodic, but their repeat lengths along the $c$ direction are incommensurate.\citep{Schwarz_2003,McMahonNelmes2004Review,McMahon_2007,janssen2006incommensurate} This coexistence of order and mismatch, mediated by host--guest interactions, makes these phases physically intriguing: they remain well ordered, yet frustrated along one direction and highly pressure tunable.\citep{McMahonNelmes2004Review,boldyreva2017breakthrough}

In Sb-II, the host framework forms a body-centered tetragonal cage structure (space group $I4/mcm$) with channels running along the $c$ axis that accommodate linear guest chains of body-centered tetragonal symmetry (space group $I4/mmm$).\citep{Schwarz_2003,McMahon_2007} Although the crystallographic structure of Sb-II has long been established, the microscopic origin of its incommensurability remains unresolved.\citep{Schwarz_2003,McMahon_2007,Coleman2018} Diffraction patterns of host--guest phases exhibit sharp Bragg reflections, often accompanied by satellite peaks,\citep{McMahon_2007,janssen2006incommensurate,McMahonNelmes2004Review} and are commonly interpreted through two related but conceptually distinct pictures. In one, they are viewed as incommensurate composites, in which the host and guest retain partially distinct periodicities and corresponding reflection sets. In the other, they are treated as incommensurately modulated structures, in which an average crystal framework is decorated by a modulation wave.\citep{janssen2006incommensurate,arakcheeva2017incommensurate,boldyreva2017breakthrough}

Both descriptions are powerful at the level of diffraction and crystallographic characterization, but they leave open a more microscopic question: what real-space structural object carries the modulation in Sb-II? Addressing this question is essential for connecting the diffraction-level description of incommensurability to the underlying local bonding, sliding energetics, and finite-temperature structural fluctuations.

Here we show that the incommensurability of Sb-II originates from a finite-density lattice of localized guest--host registry defects. Rather than a smooth harmonic modulation imposed on an average structure, the phase consists of locally commensurate 4/3 regions separated by soliton-like misregistries. This picture emerges from first-principles calculations of bonding and sliding, a Frenkel--Kontorova/sine-Gordon description of the registry phase \cite{Bak1982DevilsStaircase,BraunKivshar1998}, atomistic relaxations of rational approximants, large-scale machine-learning-potential molecular dynamics, and comparison with finite-temperature diffraction. 
\begin{figure}[!t]
  \includegraphics[width=0.98\linewidth]{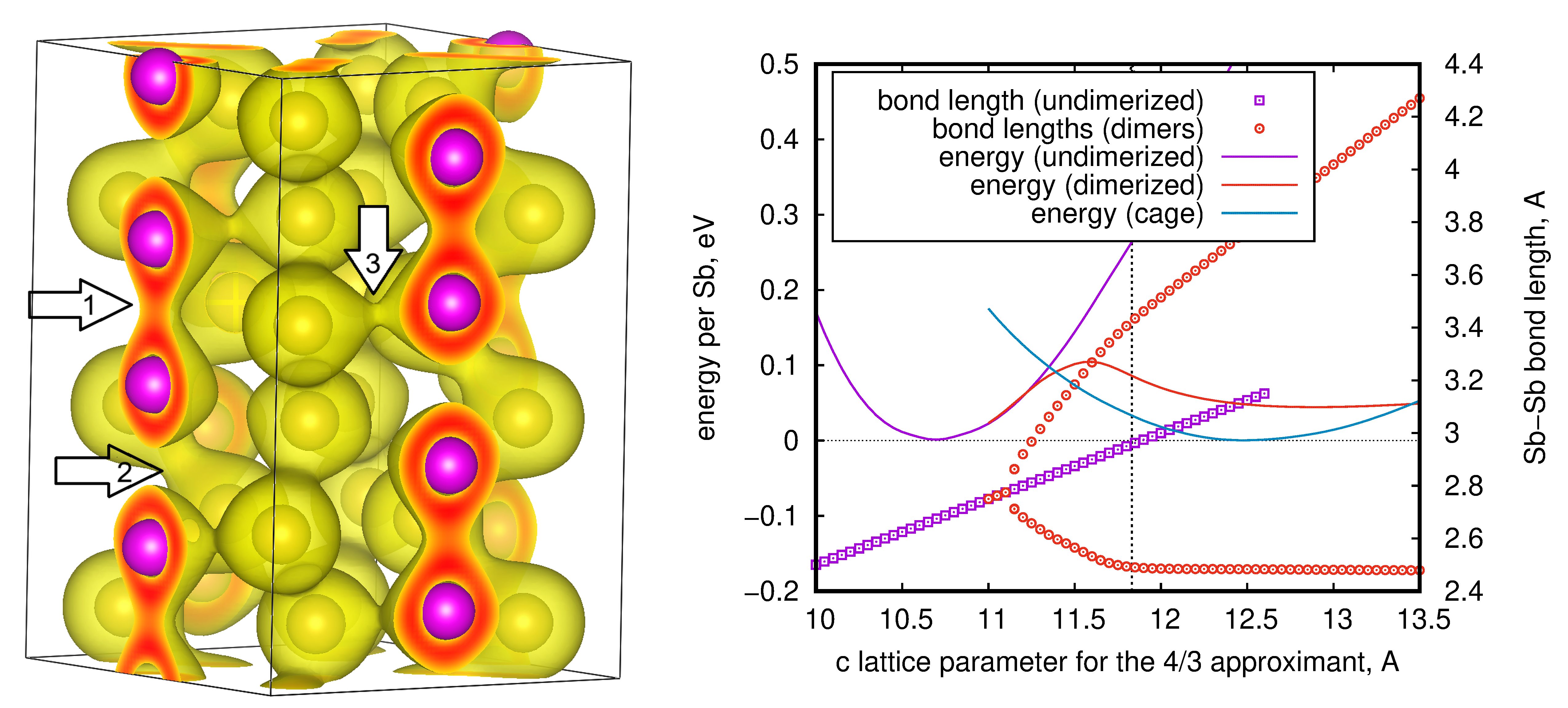}
  \caption{
  \textbf{Strong guest--host coupling and the origin of guest-chain dimerization.}
  Left: DFT-relaxed 4/3 Sb-II approximant at 10.2~GPa with charge-density isosurfaces ($n=0.037~e$/\AA$^3$); guest atoms magenta. Arrows mark (1) a guest--guest bond, (2) a host--host bond, and (3) a guest--host contact.
  Right: Energies of the isolated guest chain (undimerized and dimerized) and of the empty host cage at 10.2~GPa, per Sb atom, and the Sb--Sb distances of the chain (right axis), as functions of the $c$ lattice parameter of the 4/3 approximant. The dotted vertical line marks $c$ of the relaxed host--guest approximant. The isolated chain is undimerized at its own equilibrium ($c \approx 10.7$~\AA) but dimerizes when stretched, whereas the empty cage prefers $c \approx 12.45$~\AA; the relaxed composite lies between the two minima, so the chain is stretched and the cage compressed.
  }
  \label{fig:bonding}
\end{figure}

We first establish that the guest chains are strongly coupled to the host framework, so that the incommensurability cannot be reduced to a weak geometric mismatch. First-principles calculations were performed using the Vienna \emph{Ab initio} Simulation Package~\cite{kresse1996efficiency,kresse1996iterative} with the projector augmented-wave method \cite{kresse1999paw}. Benchmarking several Sb phases under pressure showed that the local density approximation, using the Perdew--Zunger parametrization \cite{perdew1981pz}, gives better agreement with experiment than the PBE generalized-gradient approximation \cite{perdew1996pbe}; the results below therefore use LDA unless stated otherwise (settings and benchmarks in the Supplemental Material \cite{SM}, Sec.~I). As shown in Fig.~\ref{fig:bonding}, charge-density isosurfaces reveal strong covalent bonding within the guest chains and host cages, together with appreciable charge accumulation between guest and host atoms. The guest chains therefore do not slide inside an inert channel. The energetic comparison in Fig.~\ref{fig:bonding} further shows that the isolated chain is undimerized in its ground state, ruling out a purely Peierls-like origin of the dimerization. Dimerization appears when the chain is stretched, while the cage-only minimum occurs at a longer repeat length. Thus, in the relaxed 4/3 host--guest approximant, the chain is stretched and the cage compressed; the dimerization emerges from elastic compromise and from the corrugation of the host potential under strong guest--host coupling (Supplemental Material \cite{SM}, Sec.~IV A).
\begin{figure}[!t]
    \centering
    \includegraphics[width=1.0\linewidth]{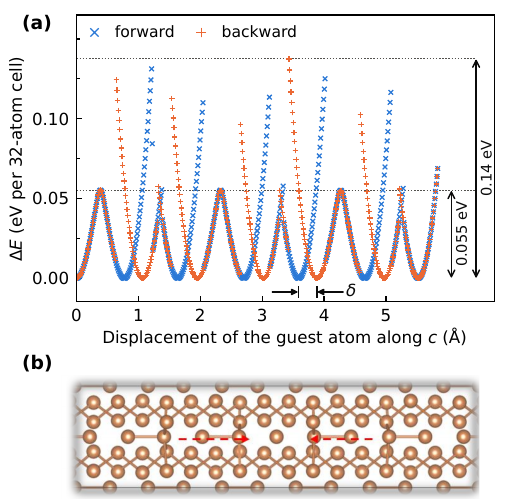}
    \caption{
    \textbf{Discrete registry states and localized phase slips.}
    (a) Sliding energy landscape (DFT, 10.2~GPa, energy per 32-atom cell) as one guest atom is displaced along the channel, forward ($\times$) and backward ($+$). Smooth maxima of 0.055~eV separate neighboring registries; the tall peaks, up to 0.14~eV, are stick--slip events at which the dragged chains jump to the next registry. $\delta$: hysteretic offset between the minima of the two sweeps.
    (b) A local phase slip interrupts the regular dimerized registry and forms a monomer-like guest--host misregistry between two commensurate segments.
    }
    \label{fig:registry}
\end{figure}
\begin{figure*}[tb]
\centering
    \includegraphics[width=0.95\textwidth]{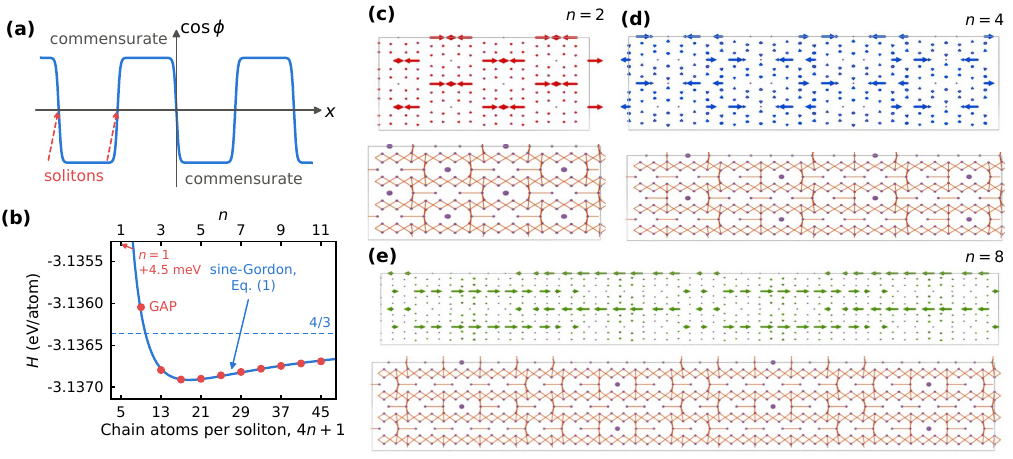}
    \caption{
    \textbf{Soliton-lattice energetics and relaxed approximants.}
    (a) Solution of Eq.~(\ref{eq:sg}) for illustrative parameters: $\cos\phi$ along the chain, with commensurate plateaus (locked 4/3 registry) separated by localized phase slips (solitons).
    (b) Enthalpy per atom (eV/atom) of the $(4n+1)/(3n+1)$ approximants as a function of the number of guest-chain atoms per soliton, $4n+1$: GAP relaxations at 10.2~GPa and 0~K (red circles, averages over ten quenches, with standard errors; $n = 1$, at +4.5~meV/atom, lies off scale) and the soliton-lattice energy of Eq.~(\ref{eq:sg}) calibrated on them (blue line; parameters in the Supplemental Material \cite{SM}, Sec.~IV); the dashed line is the commensurate 4/3 reference ($n\rightarrow\infty$). The minimum marks the preferred soliton spacing.
    (c--e) DFT-relaxed approximants with $n = 2$, 4, and 8, respectively: lower rows, the relaxed structures (host and guest atoms); upper rows, the eigenvectors of the lowest-frequency phonon mode, localized at the misregistry cores between commensurate sections.
    }
    \label{fig:soliton_lattice}
\end{figure*}

The same strong coupling produces a highly structured sliding landscape. Starting from the optimized 4/3 approximant at 10.2~GPa, we displaced one guest atom along the channel in steps of 0.001 of the cell length and relaxed the structure at fixed cell with two constraints: the displaced atom, which stays on the channel axis, and the channel coordinate of one host atom, held as a reference. All other coordinates, including the transverse coordinates of every other guest atom, were relaxed; the remaining guests of both channels follow the displaced atom, so the guest chains slide against the host. The resulting potential energy profile, Fig.~\ref{fig:registry}, is strongly corrugated, with multiple metastable minima and two characteristic barrier scales within each slip event. A dominant barrier of up to about 0.14~eV per 32-atom cell is associated with passage through host planes, where the dragged chains stick and then slip to the next registry, while a smaller barrier of about 0.055~eV per cell (about $2k_BT$ at 300~K) separates nearby metastable registries. Near each minimum the basin is approximately harmonic, with a curvature consistent with low-energy sliding modes of only a few meV. Reversing the sliding direction reveals an additional key feature: the local minima do not retrace the same registries, but are shifted by $\delta \approx 0.026$ in fractional sliding coordinate, or about 0.3~\AA\ along the channel direction. This hysteretic offset indicates that the guest--host registry is multistable and history dependent, allowing localized phase slips between nearly equivalent dimerized arrangements.

This atomistic landscape naturally motivates a coarse-grained registry-field description of the Frenkel--Kontorova (sine-Gordon) type, a standard framework for commensurate--incommensurate transitions and soliton lattices.\cite{Bak1982DevilsStaircase,BraunKivshar1998} We introduce a phase $\phi(x)$ describing the local registry of the guest-chain dimerization relative to the commensurate 4/3 reference state and write the dimerization displacement of the chain as $u_x=A\cos[Q_0x+\phi(x)]$, where $Q_0$ is the commensurate wavevector. The two values $\phi=0$ and $\pi$ represent the two degenerate dimerized registries selected by the host. The lowest-order energy functional is then
\begin{equation}
E = \frac{\kappa}{2}\int dx
\left[
\left(\partial_x \phi-\Delta Q\right)^2
- g\cos(2\phi)
\right],
\label{eq:sg}
\end{equation}
where $\kappa$ is the phase stiffness, $\Delta Q=Q-Q_0$ measures the deviation from commensurability, and $g$ is the host-induced locking strength. Minimization yields a sine-Gordon equation, equivalent to the continuum limit of a Frenkel--Kontorova model (Supplemental Material \cite{SM}, Sec.~IV A). In this description, broad plateaus of $\phi(x)$ are locally locked 4/3 regions, while sharp transitions are solitons, phase slips of $\pi$ connecting neighboring dimerized registries. A periodic array of these phase slips is therefore a real-space soliton lattice.

We use Eq.~(\ref{eq:sg}) as a semi-empirical, coarse-grained description rather than deriving it from the DFT energetics: calibrated on the GAP enthalpies, it reproduces their dependence on the soliton spacing [Fig.~\ref{fig:soliton_lattice}(b)], while all quantitative results, including the preferred spacing, are taken from the atomistic calculations, which relax every coordinate in three dimensions. Eq.~(\ref{eq:sg}) keeps only the lowest harmonic of the registry energy. This is adequate here: the higher harmonics of the corrugated landscape cannot be extracted reliably from the hysteretic landscape of Fig.~\ref{fig:registry}(a), and they change only the shape and energy of a soliton core, not the counting that yields the approximant series below. The stiffness and natural guest spacing obtained in this way agree with independent atomistic estimates, and so does the soliton width once the cell shape is relaxed (Supplemental Material \cite{SM}, Sec.~IV).

The soliton model immediately explains the rational approximants relevant to Sb-II. Inserting one phase slip every $n$ commensurate units changes the average guest--host ratio to
\[
\frac{4n+1}{3n+1},
\]
with the defect-free 4/3 structure recovered as $n\rightarrow\infty$, a result that follows from counting alone and holds for any values of the model parameters. The model also rationalizes a non-monotonic dependence of the enthalpy on the soliton spacing: closely spaced solitons repel strongly, while dilute solitons approach the commensurate reference energy. As summarized in Fig.~\ref{fig:soliton_lattice}, this competition creates a preferred finite soliton density. Direct DFT relaxations of the $(4n+1)/(3n+1)$ approximants, from 42 atoms for $n=1$ to 266 atoms for $n=8$, confirm their structure [Fig.~\ref{fig:soliton_lattice}(c--e)], but the enthalpy differences between them, mostly below 1~meV/atom, are smaller than the uncertainty of the Brillouin-zone sampling for cells of different lengths; DFT alone therefore does not resolve the preferred spacing. To resolve it, we used an independently constructed Gaussian approximation potential (GAP) \cite{Bartok2010,Bartok2013,SbGAP2026} that was agnostic to the soliton-lattice construction in its training: no configuration of approximant size is in its training set (Supplemental Material \cite{SM}, Sec.~II). The GAP calculations nevertheless recover both ingredients of the model [Fig.~\ref{fig:soliton_lattice}(b)]. Every approximant with $n\ge3$ lies below the commensurate 4/3 reference: solitons lower the enthalpy, so the commensurate structure is unstable against their formation. Closely spaced solitons repel, and the competition places the minimum at an intermediate soliton spacing, $n\approx4.4$ at 0~K and 10.2~GPa, near the experimental value $n\approx4.5$ (300~K) \cite{McMahon_2007}.  It should be mentioned that the exact position of the minimum depends sensitively on the elastic response of host and guest and on the misfit between them, and is not a prediction (Supplemental Material \cite{SM}, Sec.~IV F). The relaxed structures consist of extended commensurate sections separated by narrow misregistry cores. High-precision phonon and vibrational calculations further show that these approximants are dynamically stable, with no unstable modes and low-energy phason-like excitations localized near the soliton cores.

\begin{figure}[tb] 
\centering 
\includegraphics[width=\linewidth]{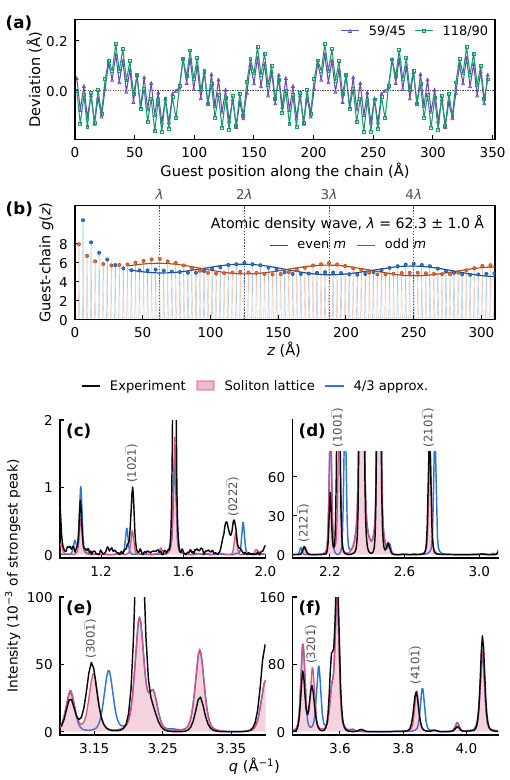} 
\caption{\textbf{Soliton lattice in large-scale ML-potential simulations, and diffraction.} (a) Guest-chain deviations from a uniformly spaced reference in the relaxed 59/45 (two periods) and 118/90 approximants; the 118/90 chain is translated along the chain to align its kinks with those of 59/45. (b) Atomic density wave in the guest-chain pair-correlation function $g(z)$ of the 21/16 approximant at 300~K: peak heights of even (blue) and odd (orange) neighbor order $m$ (dots) and a joint fit with a single wavelength $\lambda$ (lines); the two parities exchange at every soliton spacing. (c--f) One-dimensional x-ray diffraction at 10.2~GPa: experiment \cite{McMahon_2007}, the soliton lattice (21/16, shaded) and the commensurate 4/3 approximant (blue lines), after the same background subtraction, each normalized to its strongest reflection, $(2\,1\,1\,0)$. Labels mark the guest and modulation reflections; the soliton lattice reproduces all of them, the 4/3 approximant none.}
\label{fig:finiteT_xrd} \end{figure}

The same soliton physics persists in long-period structures close to the experimental host--guest ratio. We therefore relaxed the long-period approximants 59/45 and 118/90 with the machine-learning potential and performed molecular-dynamics simulations of the 21/16 approximant in supercells of more than $10^6$ atoms (geometries and simulation protocol in the Supplemental Material \cite{SM}, Sec.~III). Figure~\ref{fig:finiteT_xrd} summarizes the resulting real- and reciprocal-space signatures. To expose the underlying registry pattern, Fig.~\ref{fig:finiteT_xrd}(a) plots the deviation of guest-chain atom positions from a uniformly spaced reference. The 59/45 and 118/90 structures display the same sequence of kink-like distortions, showing that the phase-slip pattern is not an artifact of a particular approximant. These kinks are the large-scale realization of the localized guest--host misregistry defects predicted by the Frenkel--Kontorova description, spaced by the competition between their formation enthalpy and their repulsion.

At finite temperature, the low energy of the soliton modes causes strong broadening. In the large-scale ML-potential molecular dynamics at 300~K, the instantaneous soliton cores become diffuse and partially disordered, so localized defects are no longer sharply resolved in individual configurations. Their statistical signature, however, remains well defined: an \emph{atomic density wave} along the guest chains. It is resolved in the guest-chain pair-correlation function $g(z)$ of Fig.~\ref{fig:finiteT_xrd}(b), whose peaks belong to neighbors of order $m$ along a chain (Supplemental Material \cite{SM}, Sec.~III E). The heights of the even-$m$ and odd-$m$ peaks oscillate in antiphase and exchange at every $\lambda = 62.3 \pm 1.0$~\AA, the soliton spacing at this composition ($\gamma = 1.3125$, $n = 5$). This exchange identifies the atomic density wave as the soliton lattice itself. Each soliton is a $\pi$ slip of the dimerization that carries an odd number of guests, so pairs separated by an odd number of solitons see the opposite dimer parity; a sinusoidal displacement wave superposed on a uniformly dimerized chain gives no such exchange. Thermal fluctuations smear the localized phase slips, and their configurational average is a smooth atomic density wave with the period of the soliton lattice.

The same ML-potential molecular-dynamics configurations provide a stringent reciprocal-space test through the one-dimensional x-ray diffraction pattern of Sb-II \cite{McMahon_2007} (calculation details in the Supplemental Material \cite{SM}, Sec.~V). We compared the simulated pattern of the soliton-modulated structure with that of the simple commensurate 4/3 approximant and with experiment at 10.2~GPa, focusing primarily on peak positions because measured intensities are sensitive to sample texture and geometry. As shown in Fig.~\ref{fig:finiteT_xrd}(c--f), both structures reproduce the host reflections, but only the soliton lattice places the guest and modulation reflections at their measured positions: all eight labeled reflections lie within 0.01~\AA$^{-1}$, among them the guest reflection $(100\bar{1})$, the modulation reflection $(212\bar{1})$ and the second-order reflection $(022\bar{2})$, whereas the 4/3 approximant shifts them by 0.02--0.05~\AA$^{-1}$ or lacks them. Thus, the reciprocal-space modulation traditionally described in superspace emerges naturally from the same large-scale MD structure whose real-space correlations reveal a thermally broadened soliton lattice.

Taken together, these results support a real-space interpretation of Sb-II as a defect-mediated discommensurate phase. Rather than viewing the incommensurability as a simple geometric mismatch between weakly coupled periodic subsystems, as an incommensurate composite, or as an otherwise unexplained modulation imposed on an average structure \cite{Schwarz_2003,McMahon_2007,janssen2006incommensurate,arakcheeva2017incommensurate,vansmaalen2007incommensurate}, our calculations show that it can be understood microscopically as a localized guest--host soliton lattice built on a commensurate 4/3 reference state \cite{Bak1982DevilsStaircase,BraunKivshar1998}. In this picture, the ideal low-temperature structure is a periodic array of localized phase slips separating commensurate regions, whereas at finite temperature the broadened and fluctuating soliton cores generate the observed atomic density wave, which is therefore not a distinct competing object, but the statistical real-space signature of a thermally fluctuating soliton lattice \cite{Bak1982DevilsStaircase,BraunKivshar1998}. This also clarifies the relation between atomistic structure and diffraction: the superspace modulation used to describe Sb-II crystallographically \cite{janssen2006incommensurate,vansmaalen2007incommensurate,boldyreva2017breakthrough} can be viewed as the reciprocal-space manifestation of an underlying ensemble of nonlinear registry defects whose density and spacing determine the modulation period. More broadly, the soliton picture can be validated in other host--guest phases. The Frenkel--Kontorova description requires guest and host periods close to, but not at, a commensurate ratio at which the registry energy has a clear minimum (4/3 for Sb-II, with the dimerized guest chain locked to the host), and an effectively one-dimensional registry coordinate, as in channel structures. Whether the misfit then forms a soliton lattice follows from the general Frenkel--Kontorova criteria, each of which can be checked atomistically: the locking must be strong compared with the elastic cost of the misfit, so that relaxed approximants show phase slips narrow compared with their spacing (for weak locking the same equation gives a nearly sinusoidal modulation); the approximant enthalpies must fall below the commensurate reference and select a finite slip density; and creating a pair of opposite slips must cost much more than the thermal energy, so that the slip order survives at finite temperature, as the parity exchange of $g(z)$ shows here. Sb-II meets these conditions; Bi-III and other host--guest phases \cite{McMahonNelmes2004Review,McMahon_2007,Coleman2018,boldyreva2017breakthrough} can be tested in the same way.

\begin{acknowledgments}
The authors are grateful to Prof.\ M. I. McMahon for providing the high-pressure X-ray diffraction data on Sb. This project was supported by resources provided by the Office of Research Computing at George Mason University and funded in part by grants from the National Science Foundation (Award Number 2018631).
\end{acknowledgments}

 The data that support the findings of this article are openly available at DOI \href{https://doi.org/10.5281/zenodo.23514945}{10.5281/zenodo.23514945}.

\bibliography{SbII}

\end{document}

\title{Supplemental Material for ``Soliton-lattice origin of incommensurability in high-pressure antimony''}
\author{M. Smirman}
\affiliation{Department of Physics and Astronomy, George Mason University, Fairfax, VA 22030, USA}
\author{K. Sun}
\affiliation{Department of Physics, University of Michigan, Ann Arbor, Michigan 48109, USA}
\author{H. Sheng}
\affiliation{Department of Physics and Astronomy, George Mason University, Fairfax, VA 22030, USA}
\author{I. I. Mazin}
\affiliation{Department of Physics and Astronomy, George Mason University, Fairfax, VA 22030, USA}
\affiliation{Quantum Science and Engineering Center, George Mason University, Fairfax, VA 22030, USA}
\maketitle
\setcounter{tocdepth}{2}
\tableofcontents

\section{Density-functional calculations and the choice of functional}\label{sec:dft}

\subsection{Computational package}
All density-functional (DFT) calculations used VASP~6.3.2~\cite{kresse1996iterative},
the projector augmented-wave method~\cite{sm-Bloechl1994,kresse1999paw}, the standard Sb potential with five valence electrons ($5s^2 5p^3$), the local density approximation (LDA;
Ceperley--Alder data~\cite{sm-Ceperley1980} parameterized by Perdew and Zunger~\cite{perdew1981pz}),
and a plane-wave cutoff of 500~eV. Spin--orbit coupling (SOC) was omitted. Our LDA calculations of A7 Sb with and without SOC show that it increases the volume by 0.2--0.4\%, changes the internal coordinate by $(1$--$3)\times10^{-4}$,
and shifts the zone-center phonon frequencies by at most 5~cm$^{-1}$. These changes are 10 to 30 times smaller than the differences between LDA and PBE in volume (8\%) and internal coordinate
($3\times10^{-3}$), and four times smaller than their difference in the $E_g$ frequency at the relaxed cells (20~cm$^{-1}$; Sec.~\ref{sec:functional}).

\subsection{Choice of functional}\label{sec:functional}
To benchmark against the PBE generalized-gradient approximation~\cite{perdew1996pbe}, we used the standard Sb PBE and LDA potentials (i.e., not the hard-core versions). A7 Sb was computed in its two-atom cell with a $24\times24\times24$ $k$-point mesh and Methfessel--Paxton smearing~\cite{sm-Methfessel1989} of 0.05~eV; Sb-II (32-atom 4/3 approximant) and bcc Sb were relaxed at the pressures of
the measurements (atomic positions and cell) with a $k$-point spacing of 0.10~\AA$^{-1}$ and Methfessel--Paxton smearing of 0.1~eV.

\noindent We chose the LDA over the PBE generalized-gradient approximation because it reproduces the measured structures of antimony more closely, both for ambient-pressure A7 (Sb-I) and for the high-pressure phases (Table~\ref{tab:func}). PBE overestimates the volume of A7 by 5.1\% (LDA: $-3.2$\%) and the in-plane lattice parameter by 1.6\% (LDA:
$-0.17$\%). With the cell fixed at the experimental lattice, LDA relaxes the A7 internal coordinate to $u = 0.2345$ and PBE to $0.2355$, against $0.2335$ measured: LDA has half the PBE deviation in the Peierls distortion. In the high-pressure phases the PBE volume error decreases but remains larger than the LDA error and of opposite sign: $+2.8\%$ (PBE) against $-1.9\%$ (LDA) for Sb-II at 10.2~GPa, and $+2.4\%$ against $-1.0\%$ for bcc Sb-III at 28--30~GPa. All volumes are static (0~K, no zero-point
motion) and are compared with room-temperature data; zero-point and thermal expansion, of order 1\% in volume, move the LDA values toward experiment and the PBE values away from it. Under pressure (0--8~GPa), PBE gives the A7 internal coordinate closer to experiment at the same pressure (rms deviation of $u$ 0.0018, against 0.0054 for LDA), but only because its volume is too large: compared at the same volume, both functionals slightly underestimate the Peierls distortion, LDA less than PBE (rms deviation 0.0014 against 0.0023; experimental $u(V)$ of Refs.~\cite{sm-Schiferl1981,sm-Degtyareva2004HPR}, as compiled in Ref.~\cite{wang2006effect}). We do not base oru choice on zone-center phonon frequencies: at the experimental lattice with converged Brillouin-zone sampling the two
functionals give the A7 $A_{1g}$ and $E_g$ frequencies within 1--2~cm$^{-1}$ of each other.

\emph{Limitations of the LDA.} The LDA slightly underestimates the volume at a given pressure (by 1--3\%,
Table~\ref{tab:func}), so it reaches the volume at which the Peierls distortion of A7 vanishes at a lower pressure than experiment: in our static LDA calculations A7 closes to simple cubic between 6 and 7~GPa, whereas experimentally it remains distorted up to the
transition to the host--guest phase near 8.5~GPa. At fixed volume the LDA also slightly underestimates the distortion itself (see above). For the same reason, pressures of transitions between phases of antimony obtained from static LDA enthalpies turn out to be lower than observed, without zero-point and thermal contributions included; we therefore refrain from quoting transition pressures from the LDA results. The results of this Letter are not affected.

\begin{table}[h]
\caption{LDA and PBE against experiment for Sb phases (static, 0~K). Deviations from experiment in parentheses.
A7: experiment at 298~K~\cite{Barrett1963Sb}; $u$ at the experimental
lattice with the atoms relaxed. bcc: experimental equation of state of Ref.~\cite{Degtyareva_2004}. Sb-II: experimental incommensurate structure of Ref.~\cite{McMahon_2007},
against the 32-atom 4/3 approximant ($c = 3c_\mathrm{H} = 4c_\mathrm{G}$), whose guest density exceeds the
experimental one by 1.7\%, i.e.\ by 0.4\% in atoms per volume.}
\label{tab:func}
\begin{ruledtabular}
\begin{tabular}{llccc}
Phase & Quantity & LDA & PBE & Expt. \\
\hline
A7, 0~GPa       & $V$ (\AA$^3$/atom)        & 29.24 ($-3.2$\%) & 31.76 ($+5.1$\%) & 30.21 \\
                & $a_\mathrm{hex}$ (\AA)    & 4.301 ($-0.17$\%) & 4.378 ($+1.6$\%) & 4.308 \\
                & $c_\mathrm{hex}$ (\AA)    & 10.95 ($-2.9$\%) & 11.48 ($+1.8$\%) & 11.27 \\
                & $u$ at expt.\ lattice     & 0.2345 & 0.2355 & 0.2335 \\
bcc, 28~GPa     & $V$ (\AA$^3$/atom)        & 20.33 ($-1.0$\%) & 21.03 ($+2.4$\%) & 20.54 \\
bcc, 30~GPa     & $V$ (\AA$^3$/atom)        & 20.12 ($-0.9$\%) & 20.79 ($+2.4$\%) & 20.31 \\
Sb-II, 10.2~GPa & $V$ (\AA$^3$/atom)        & 23.50 ($-1.9$\%) & 24.62 ($+2.8$\%) & 23.94 \\
                & $a$ (\AA)                & 8.005 ($-0.8$\%) & 8.130 ($+0.8$\%) & 8.068 \\
\end{tabular}
\end{ruledtabular}
\end{table}

\subsection{Charge densities, sliding landscape and approximant relaxations}
The calculations for Figs.~1--3 of the Letter used Gaussian smearing of 0.03~eV and an electronic convergence of $10^{-8}$~eV, with ionic relaxations converged to $10^{-7}$~eV. The 4/3 approximant (Fig.~1) and the
$(4n+1)/(3n+1)$ approximants (Fig.~3) were relaxed at 10.2~GPa (atomic positions and cell) with a $k$-point spacing of 0.15~\AA$^{-1}$. The sliding landscape (Fig.~2) was computed at fixed cell, that of the 4/3 approximant relaxed at
10.2~GPa ($a = 8.036$~\AA, $c = 11.674$~\AA, 32 atoms), with a $\Gamma$-centered $4\times4\times4$ $k$-point mesh; its constraints are described in Sec.~\ref{sec:md}.

\subsection{Phonons of the approximants}
The harmonic phonons of the approximants [Fig.~3(c--e) of the Letter] were computed by density-functional perturbation theory~\cite{sm-Baroni2001} on the DFT-relaxed structures, without symmetry, with high numerical precision, Methfessel--Paxton smearing of 0.2~eV, an electronic convergence of $10^{-8}$~eV, and the $k$-point spacing of the relaxations, 0.15~\AA$^{-1}$.

\subsection{Training data of the potential}
The DFT energies, forces and virials of the training set of the interatomic potential (Sec.~\ref{sec:gap}) were computed with a smearing width of 0.05~eV and a $k$-point spacing of 0.15~\AA$^{-1}$.

\clearpage
\section{Gaussian approximation potential (GAP-S\lowercase{b})}\label{sec:gap}

\subsection{Model}
The GAP-Sb potential, a Gaussian approximation potential (GAP)~\cite{Bartok2010,sm-Deringer2021}, combines a two-body term and a many-body smooth-overlap-of-atomic-positions (SOAP) term~\cite{sm-Bartok2013PRB}. It was fitted with the \texttt{gap\_fit} program of the QUIP package~\cite{sm-QUIP} with the settings listed in Table~\ref{tab:gapfit}. A constant per-atom energy offset, $e_0 = -4.00728$~eV, is subtracted before fitting.

\begin{table}[h]
\caption{GAP-Sb fit settings (QUIP/GAP \texttt{gap\_fit}).}\label{tab:gapfit}
\begin{ruledtabular}
\begin{tabular}{lp{0.72\textwidth}}
Two-body term & cutoff 5.0~\AA; 100 sparse points (uniform); squared-exponential kernel, $\theta = 1.0$; $\delta = 2.0$~eV \\
SOAP term & cutoff 5.0~\AA, transition width 1.0~\AA; $n_\mathrm{max} = 12$, $l_\mathrm{max} = 4$; atomic width 0.5~\AA; radial decay $-0.5$; dot-product kernel, $\zeta = 4$; 9000 sparse environments (CUR~\cite{sm-Mahoney2009}); $\delta = 0.2$~eV \\
Energy offset & $e_0 = -4.00728$~eV per atom \\
Core term & the fit file specifies a Glue core potential (\texttt{r6\_innercut.xml}) that has no Sb parameters and contributes nothing; listed for reproducibility only \\
Regularization ($\sigma$) & energy 1~meV, forces 0.01~eV/\AA, virials 0.05~eV (per configuration) \\
Weighted subset & 150 Sb-II/Sb-IV configurations (4/3 approximant minima and images along the path between the two registries): energy $\sigma = 0.2$~meV \\
Sparse jitter & $10^{-8}$ \\
\end{tabular}
\end{ruledtabular}
\end{table}

\subsection{Training set}
The training set contains 4517 LDA configurations with 170\,383 atoms (Table~\ref{tab:train}). It covers crystalline
prototypes: 29 elemental structures from the Open Quantum Materials Database (OQMD)~\cite{sm-Saal2013,sm-Kirklin2015} and six known structures of Sb (A7 in its 2-atom primitive cell and a 54-atom cell, simple cubic, bcc, the 4/3 Sb-II approximant, and the black-phosphorus A17 type) from Pearson's Crystal Data~\cite{sm-PCD2020}. Each structure enters at ten isotropic scalings of its lattice vectors, by factors of 0.85 to 1.10, both unperturbed and with all atoms displaced randomly from their sites (Gaussian displacements with standard deviations of 0.01, 0.02, 0.04, 0.06 and 0.08~\AA), so that the set samples compression, expansion, and the forces of distorted crystals. The training set further contains configurations collected by active learning during stochastic-surface-walking (SSW) structure
searches~\cite{sm-Shang2013}: during our iterative training steps, SSW searches were run with the current version of the potential, and the structures they visited were recomputed with DFT and added to the training set for the next fit.

These searches, with their minima re-relaxed with GAP-Sb, also produced a candidate atomic structure for the host--guest phase Sb-IV, which will be reported in a separate paper~\cite{sm-SmirmanSbIV}. The training set also contains snapshots of ab initio molecular dynamics (AIMD) trajectories: liquids at several densities and temperatures, melt--quench runs, and Sb-I, Sb-II and simple-cubic crystals at finite temperature. It further includes surfaces and slabs, clusters and dimers, scans of the A7 internal coordinate, and the Sb-II/Sb-IV configurations that receive the higher weight. No configuration in the training set has the size of a soliton-lattice approximant [$32n+10$ atoms for $(4n+1)/(3n+1)$: 42 for $n = 1$ to 266 for $n = 8$], so the soliton-lattice results are a test of the potential, not an input to it.

\begin{table}[h]
\caption{Composition of the GAP-Sb training set and fitting errors on it (energy per atom and virial per atom:
root-mean-square; forces: root-mean-square per Cartesian component). Distinct configurations in parentheses.}
\label{tab:train}
\begin{ruledtabular}
\begin{tabular}{lrrccc}
Category & Configs & Atoms & $E$ (\meva) & $F$ (eV/\AA) & Virial (\meva) \\
\hline
Crystal prototypes                  & 3081  & 28\,479 & 10.3 & 0.113 & 109 \\
Active learning (SSW searches)      &  433   &  9\,132 &  4.3 & 0.052 &  36 \\
Liquid and melt--quench MD          &  290   & 72\,500 &  5.9 & 0.121 &  29 \\
Sb-I, Sb-II, sc at finite $T$       &  240        & 28\,400 &  6.7 & 0.139 &  40 \\
Clusters, dimers, surfaces          &  131        &  3\,050 & 26.8 & 0.170 & 121 \\
A7 scans and static sets            &  192   & 12\,822 &  3.3 & 0.065 &  31 \\
Sb-II/Sb-IV set (weighted)          &  150        & 16\,000 & 0.69 & 0.028 &  25 \\
\hline
Total                                           & 4517  & 170\,383 & 10.0 & 0.113 &  94 \\
\end{tabular}
\end{ruledtabular}
\end{table}

\subsection{Validation}
\begin{figure*}[p]
\includegraphics[width=\textwidth]{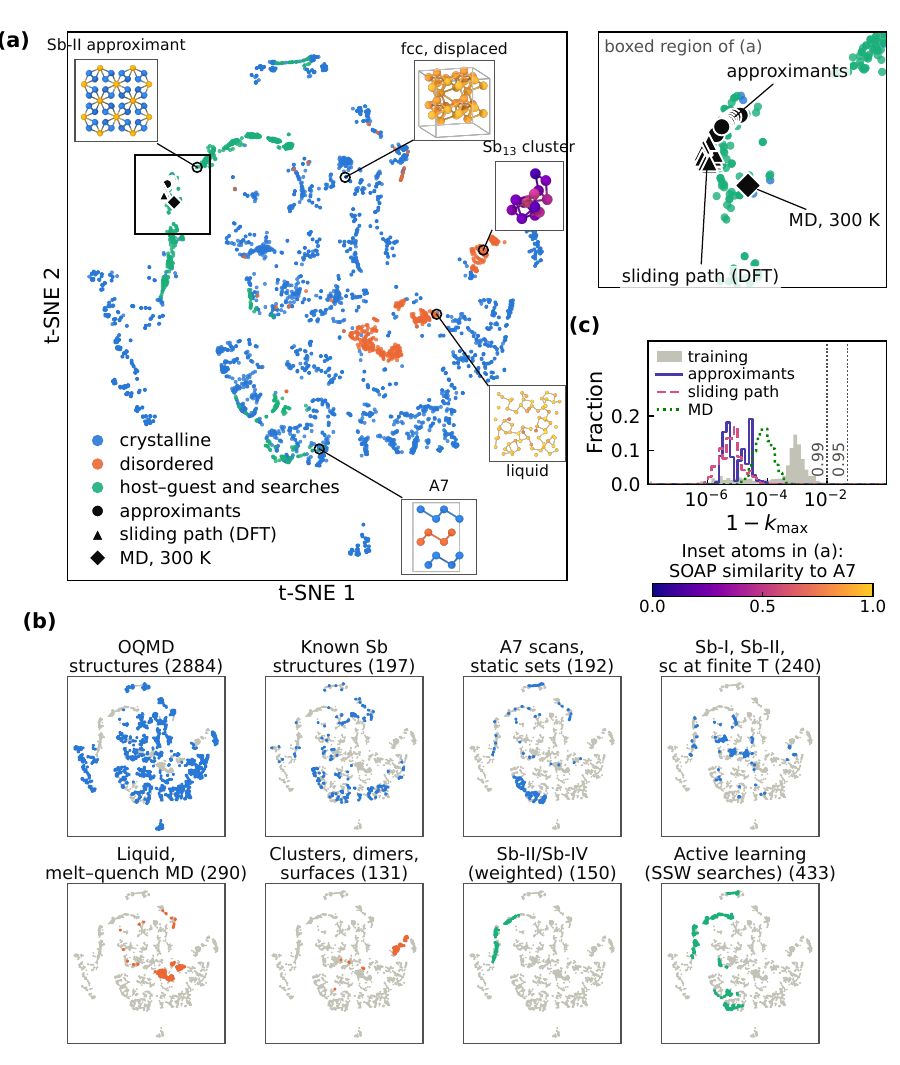}
\caption{The GAP-Sb training set in descriptor space. (a)~t-SNE map~\cite{sm-vanderMaaten2008} (perplexity 50; computed
with ASAP~\cite{sm-Cheng2020}) of the per-configuration SOAP descriptors of the 4517 training configurations, colored by group: crystalline (blue), disordered (orange), and host--guest structures and stochastic-surface-walking (SSW) searches (green). Black symbols mark the configurations used in the Letter (approximants, the DFT sliding path, and 300~K molecular dynamics); the upper right panel enlarges the boxed region. Insets in (a) show representative training structures; the atoms of the displaced fcc, Sb$_{13}$-cluster and liquid insets are colored by their SOAP similarity to A7 (color bar below (c)); A7 is colored by bilayer and Sb-II by host (blue) and guest
(amber).
(b)~The eight types of training configurations on the same map (number of configurations in parentheses).
(c)~Coverage: distribution of $1 - k_\mathrm{max}$, where $k_\mathrm{max}$ is the largest SOAP kernel similarity of an
atom to any training atom, for the Letter's configurations and, for reference, for training atoms with their own
configuration left out (gray). Every atom of the Letter's configurations has $k_\mathrm{max} > 0.999$.}
\label{fig:train}
\end{figure*}

\begin{figure*}[p]
\includegraphics[width=\textwidth]{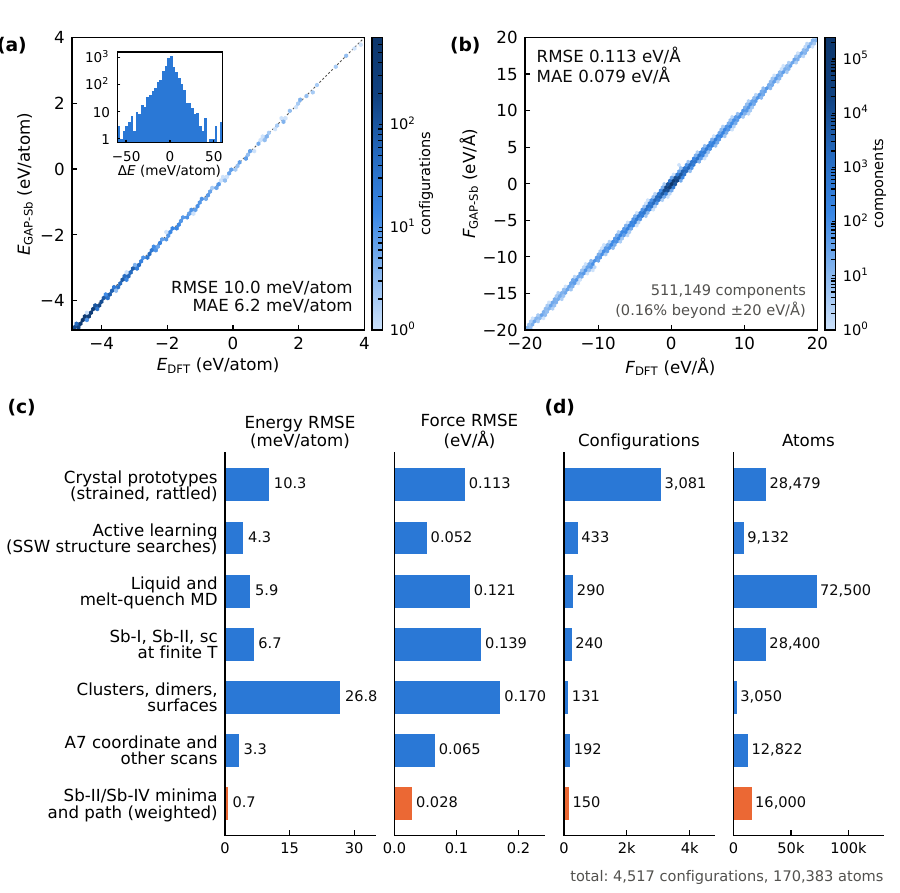}
\caption{Fit quality of GAP-Sb on its training set (4517 configurations, 170\,383 atoms). (a)~Energy per atom from GAP-Sb against DFT for every training configuration, shown as a
two-dimensional histogram (hexagonal cells colored by the number of configurations, logarithmic scale); inset, distribution of the error. Root-mean-square (RMSE) and mean absolute errors: 10.0 and 6.2~meV/atom. (b)~Every Cartesian force component (511\,149) from GAP-Sb against DFT; RMSE 0.113~eV/\AA, mean absolute error 0.079~eV/\AA; the 0.16\% of
components above 20~eV/\AA, from strongly compressed or displaced cells, lie outside the plotted range. (c)~Energy and force RMSE for each type of training configuration; the Sb-II/Sb-IV minima and the configurations along the path between them (orange) were fitted with a higher weight and are reproduced to 0.7~meV/atom and 0.028~eV/\AA.
(d)~Number of configurations and of atoms of each type.}
\label{fig:fit}
\end{figure*}

\begin{figure*}[p]
\includegraphics[width=\textwidth]{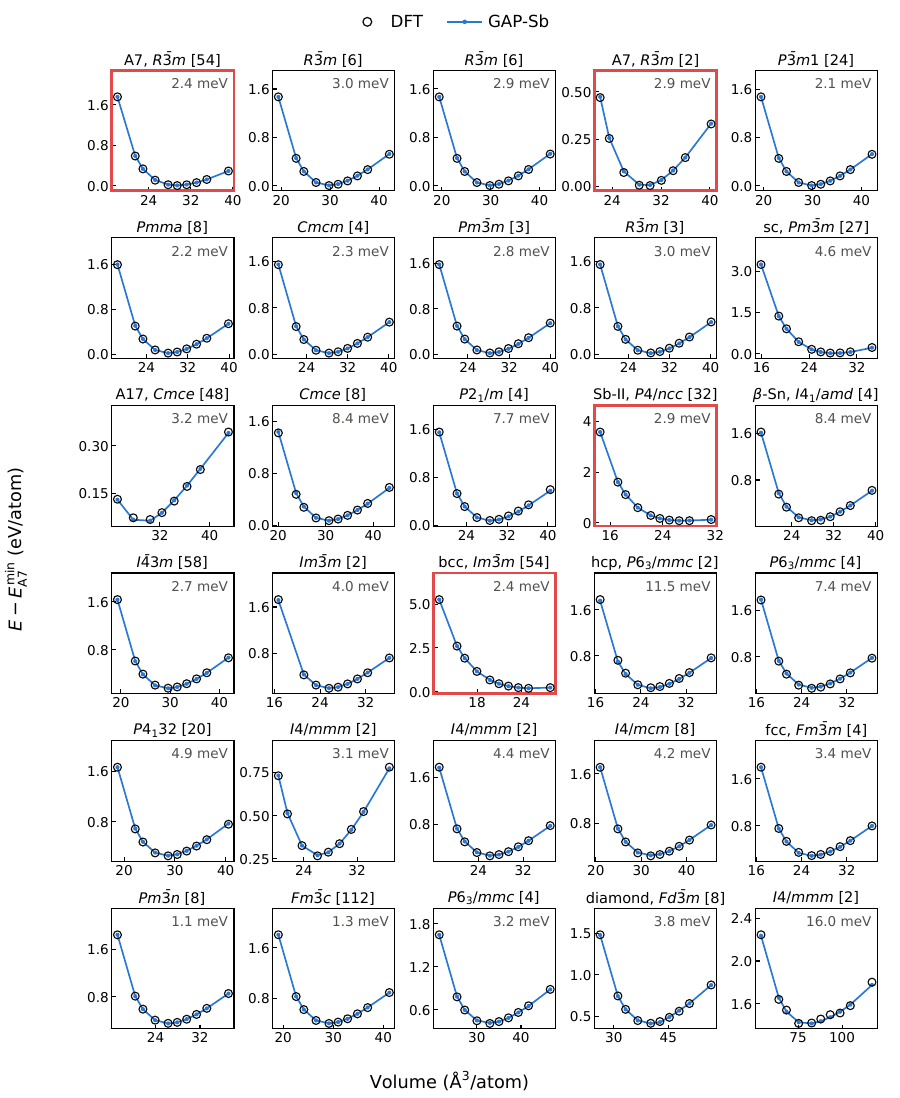}
\caption{Energy--volume curves of the crystal prototypes in the training set: energy per atom relative to the minimum of A7 against volume per atom for the unperturbed affine transformations (lattice vectors scaled by 0.85--1.10) of 30 distinct prototypes, DFT (open circles) and GAP-Sb on the same structures (filled circles; lines are guides). Twenty-four structures are from the OQMD~\cite{sm-Saal2013,sm-Kirklin2015} and six from Pearson's Crystal Data (A7 in its 54-atom and 2-atom cells, simple cubic, bcc, A17, and the
32-atom 4/3 Sb-II approximant); red frames mark the phases of Sb observed in experiment (A7, Sb-II and bcc). Scans of duplicate OQMD structures are merged. Panels are ordered by the DFT minimum energy and labeled by name and space group, with the number of atoms per cell in brackets; the number in each panel is the GAP-Sb energy RMSE over that scan. Over all 295 structures the RMSE is 5.5~meV/atom. The structures belong to the training set, so the figure shows the quality of the fit across structure types; tests on configurations outside the training set are shown in Fig.~\ref{fig:valid}.}
\label{fig:eos}
\end{figure*}

\begin{figure*}[p]
\includegraphics[width=\textwidth]{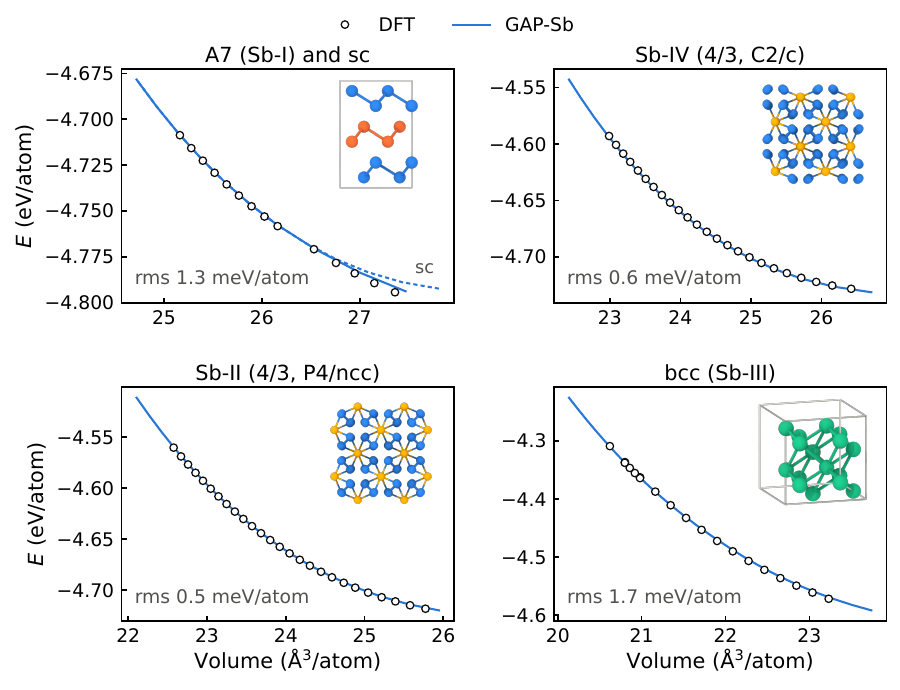}
\caption{Energy--volume curves of the phases of compressed Sb: DFT relaxations of 128-atom cells at fixed pressure (open circles; A7 and sc 3.5--10~GPa, Sb-IV 1.5--13~GPa, Sb-II 2.5--15~GPa, bcc 8.4--25~GPa) and GAP-Sb relaxations
at 0--30~GPa (lines; dashed, simple cubic). The number in each panel is the root-mean-square difference of the GAP-Sb and DFT energies at equal volume: 1.3 (A7/sc), 0.6 (Sb-IV), 0.5 (Sb-II) and 1.7~meV/atom (bcc). The DFT structures belong to the training set. Insets: the structures (A7 bilayers in alternating blue and orange; bcc green; Sb-IV and Sb-II viewed down the channels, host blue, guest amber). Note: the atomic structure of Sb-IV shown here will be reported elsewhere (Sec.~II B).}
\label{fig:eosph}
\end{figure*}

\begin{figure*}[p]
\includegraphics[width=\textwidth]{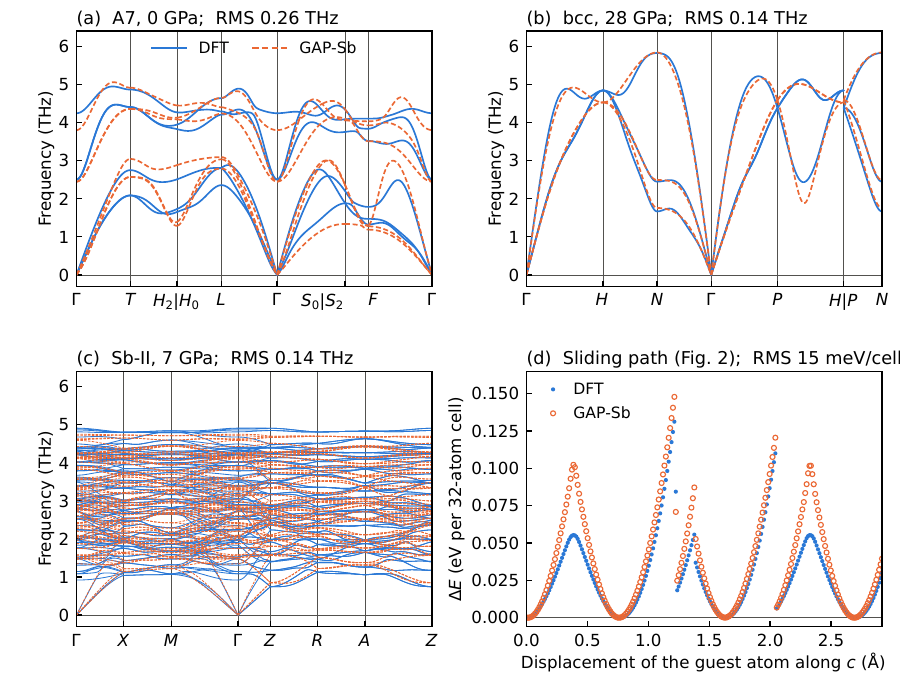}
\caption{GAP-Sb against DFT on structures it was not fitted to. (a)--(c)~Harmonic phonon dispersions computed by finite displacements on identical 128-atom supercells and displacement sets, DFT (solid) and GAP-Sb (dashed):
(a)~A7 at 0~GPa, (b)~bcc at 28~GPa, (c)~the 4/3 Sb-II approximant at 7~GPa; the number in each title is the root-mean-square difference of all branch frequencies along the path. No displaced supercell is in the training set; the undisplaced Sb-II supercell is, so (c) tests the force response around a training structure. (d)~Energy along the forward sliding sweep of Fig.~2 of the Letter (32-atom cell, 10.2~GPa), DFT (filled) and GAP-Sb single points on the DFT geometries (open), relative to the first point; no sliding configuration is in the training set.}
\label{fig:valid}
\end{figure*}

Figure~\ref{fig:train} shows where the training configurations lie in descriptor space and that the atomic environments of the structures simulated in the Letter are covered by them. Figure~\ref{fig:fit} compares GAP-Sb with the DFT training data, and Figs.~\ref{fig:eos} and \ref{fig:eosph} the energy--volume curves of the crystal prototypes and of the phases of compressed Sb. The tests in
Fig.~\ref{fig:valid} use DFT data that are not part of the training set. On the training data, GAP-Sb reproduces the DFT energies to 10.0~meV/atom and the forces to 0.11~eV/\AA\ (root-mean-square), and the weighted Sb-II/Sb-IV set to 0.7~meV/atom and 0.03~eV/\AA; the energy--volume curves of the 30 crystal prototypes agree to 5.5~meV/atom and those of A7, Sb-IV, Sb-II and bcc to 0.5--1.7~meV/atom. The harmonic phonon dispersions of A7, bcc and Sb-II, computed on identical supercells and displacements, agree with DFT to 0.26, 0.14 and 0.14~THz. Along the DFT sliding path of Fig.~2 of the Letter, GAP-Sb reproduces the positions of the registry minima and of the stick--slip events, and the heights of the stick--slip peaks to within 0.02~eV per 32-atom cell, but it places the smooth maxima between neighboring registries at 0.10~eV per cell, against 0.055~eV in DFT. The registry corrugation of GAP-Sb is therefore stronger than that of DFT, which affects the energies and widths of the soliton cores; the position of the preferred soliton spacing in Fig.~3(b) carries this uncertainty.

\clearpage
\section{Atomistic simulations: geometries, supercells and molecular dynamics}\label{sec:md}
All atomistic simulations use periodic boundary conditions in three dimensions, with the host channels and guest chains along $c$. LAMMPS (2 Aug 2023)~\cite{sm-Thompson2022}
was used with the GAP potential of Sec.~\ref{sec:gap} through the QUIP interface.

\subsection{Sliding landscape (Fig.~2)}
The 32-atom 4/3 approximant (24 host and 8 guest atoms, $c = 3c_H = 4c_G$),
relaxed with DFT at 10.2~GPa ($a = 8.036$~\AA, $c = 11.674$~\AA), was kept at fixed cell. One guest atom was displaced along $c$ in steps of $0.001c$ (0.0117~\AA), 500 steps forward to $0.5c$ (5.84~\AA) and 500 steps back.
After each step the structure was relaxed with the displaced atom held fixed and one host atom fixed along $c$; all other coordinates, including the transverse coordinates of the other guests, were free. Each step started from the previous relaxed structure, so each sweep follows its
metastable branch, which gives the stick--slip events and the hysteresis of Fig.~2(a). DFT settings are given in
Sec.~\ref{sec:dft}.

\subsection{$(4n+1)/(3n+1)$ approximants (Fig.~3)}
Each cell contains one conventional host cell in the plane (two channels) and $3n+1$ host periods along $c$. In each channel, $n$ commensurate 4/3 blocks taken from the relaxed 4/3 approximant are joined by one 1/1 core, giving $2(4n+1)$ guest and $8(3n+1)$ host atoms ($32n+10$ atoms: 42 for
$n = 1$ to 266 for $n = 8$; the GAP calculations extend to $n = 11$, 362 atoms). Each ratio is the mediant of $n$ copies of 4/3 and one 1/1, and a Farey neighbor of 4/3 [$4(3n+1) - 3(4n+1) = 1$]~\cite{sm-Farey1816,sm-HardyWright2008}: the series approaches 4/3 along one
branch of the Farey tree~\cite{Bak1982DevilsStaircase}, and the experimental ratio 1.3104 lies between its members 17/13 and 21/16.
DFT relaxations used the settings of Sec.~\ref{sec:dft}. For the GAP enthalpies, each approximant and the 4/3 reference was quenched ten times: 0.2~ps of constant-energy molecular dynamics from random velocities for 100~K (a different seed each time), followed by conjugate-gradient minimization at 10.2~GPa with isotropic cell relaxation (energy and force tolerances $10^{-12}$).
The enthalpy per atom of Fig.~3(b), $(E + PV)/N$ at $P = 10.2$~GPa, is the average over the quenches, and its error bar the standard error of that average; the quenches of one cell differ by at most 0.11~meV/atom (standard deviation; below 0.003~meV/atom for $n \ge 5$). Two checks used the same starts. In $4\times4\times2$ supercells of the approximants (32 channels, two soliton periods along $c$, up to 11\,584 atoms; $4\times4\times4$ for the 4/3
reference), quenched ten times in the same way (velocities for 50~K, force tolerance $10^{-6}$~eV/\AA), the enthalpies of Fig.~3(b) are reproduced to within 0.06~meV/atom, with the minimum at $n \approx 4.4$--4.5. With the cell shape relaxed as well (free cell, otherwise the same protocol), the approximants are analyzed in Sec.~\ref{sec:fkagree}; the same supercells reproduce these free-cell enthalpies to within 0.08~meV/atom and their calibration in
Table~\ref{tab:fkcell} (minimum at $n \approx 2.9$--3.0, $c_G^{\mathrm{nat}} = 3.014$--3.016~\AA, width 3.3--3.5 guest spacings).

\subsection{Long-period approximants [Fig.~4(a)]}
The 59/45 cell (478 atoms: 45 host periods and 59 guests in each
of the two channels; $a = 8.03$~\AA, $c = 175.6$~\AA) was built with uniformly spaced guest chains, so that neither the modulation nor the misregistry cores were imposed. It was relaxed by conjugate gradients with isotropic cell relaxation at 10.2~GPa (tolerances $10^{-12}$). The
118/90 cell was built from 30 repeats of the 4/3 approximant along $c$ by placing 118 guests uniformly in each channel, i.e.\ removing two guests per channel; in the shifted 118/90 cell all guests were displaced by an arbitrary
distance along the channel before relaxation. Both were relaxed in the same way, at 10.2~GPa.

\subsection{Molecular dynamics [Fig.~4(b--f)]}
The soliton-lattice structure was simulated with LAMMPS and GAP-Sb in the 21/16 approximant ($n = 5$). Its unit cell contains 170 atoms (16 host periods along $c$ and 21 guests in each of the two channels; $c = 62.5$~\AA). The MD cell is this unit cell replicated $8 \times 40 \times 30$ times: 1\,632\,000 atoms in 640 guest chains of 630 atoms, $64.0 \times 319.8 \times 1875.7$~\AA\ after relaxation. The relaxation used conjugate gradients with isotropic cell relaxation at 10.2~GPa (tolerances $10^{-12}$); the system was then given random velocities for 100~K and integrated in the NPT ensemble at 10.2~GPa, for 5~ps at 100~K and then for 10~ps at 300~K (time step 1~fs). The commensurate 4/3 approximant (32-atom unit cell, replicated $8 \times 80 \times 55$ times: 1\,126\,400 atoms, $64.1 \times 640.7 \times 645.8$~\AA\ after relaxation) was simulated with the same protocol. Figure~4(b) uses the last 300~K configuration of the 21/16 run, and Fig.~4(c--f) the last 300~K configurations of both runs.

\subsection{Atomic density wave: guest-chain pair correlation [Fig.~4(b)]}\label{sec:gz}
The guest chains are identified in the 300~K configuration itself: the atoms are projected onto the plane perpendicular to $c$, each projected atomic column is given the area of its cell in the periodic Voronoi tessellation of the columns, and the guest columns, which have the larger cells (about 7.7 against 6.2~\AA$^2$), are separated from the host columns by a two-cluster split of the areas. This finds all 640 guest chains of 630 atoms.
The pair-correlation function counts only pairs within one chain,
\begin{equation}
g(z) = \frac{\sum_{\mathcal{C}} \sum_{i<j \in \mathcal{C}} \chi_{\Delta z}\!\left(z - |z_i - z_j|_L\right)}
            {\sum_{\mathcal{C}} N_{\mathcal{C}} (N_{\mathcal{C}} - 1)\, \Delta z / L},
\label{eq:gz}
\end{equation}
where $N_{\mathcal{C}}$ is the number of atoms in chain $\mathcal{C}$, $|\cdot|_L$ the minimum-image separation along $c$, and $\chi_{\Delta z}$ selects the bin of width $\Delta z = 0.05$~\AA; $g(z) \to 1$ for uncorrelated positions.

For an ordered chain, $g(z)$ has one sharp peak for each neighbor order $m$ (the pairs $i$, $i+m$), and the atomic density wave is the modulation of the peak heights $h_m$. In a dimerized chain, pairs of even order all have the same separation and give sharp peaks, while pairs of odd order are split by the bond alternation. Each soliton, a $\pi$ slip of the dimerization carrying an odd number of guests (21 here), exchanges these roles, so the even and odd peaks become sharp in turn: each parity oscillates with period $2\lambda$, and the two in antiphase. A joint fit to the
peak heights for $z_m > 40$~\AA,
\begin{equation}
h_m = h_0 + b\,(z_m - z_0) \pm A\cos(\pi z_m/\lambda + \phi),
\label{eq:hm}
\end{equation}
(upper sign for even, lower for odd $m$; $z_0 = 150$~\AA) gives $\lambda = 62.3 \pm 1.0$~\AA, against
62.5~\AA\ for one soliton per 21 guests. Since the mean spacing is fixed by the composition, what $g(z)$ shows is that at 300~K the dimerization still slips by $\pi$ at that spacing. A model chain with the same mean spacing, bond alternation and thermal broadening reproduces the exchange with a $\pi$ slip every 21 atoms, but not with a sinusoidal modulation of the same period and no slips, for which the even peaks are taller at every separation.

\clearpage
\section{Frenkel--Kontorova / sine-Gordon model}\label{sec:fk}
\subsection{Model and soliton-lattice solution}
Eq.~(1) of the Letter,
\begin{equation}
E = \frac{\kappa}{2}\int dx\,\left[(\partial_x\phi - \Delta Q)^2 - g\cos 2\phi\right],
\end{equation}
describes the guest chain by a phase $\phi(x)$. Measured from the sites $x_j = jc_G$ of a uniform chain with the
commensurate spacing, the displacement of guest $j$ has an alternating part, the dimerization
$A\cos[Q_0 x_j + \phi(x_j)] = A(-1)^j\cos\phi(x_j)$ with $Q_0 = \pi/c_G$, and a smooth part, the registry shift
$(c_G/3\pi)\,\phi(x_j)$ derived below. In the 4/3 approximant $\phi$ is constant, 0 or $\pi$, its two
dimer pairings [Fig.~2(a) of the Letter]. In a soliton lattice, however, $\phi$ steps from one to the other at each core, so that $\cos\phi$, which sets the
sign of the dimerization, alternates between plateaus of $+1$ and $-1$ [Fig.~3(a) of the Letter].

The three parameters of Eq.~(1) are
\begin{itemize}\setlength{\itemsep}{0pt}\setlength{\parskip}{0pt}
\item[$\bullet$] \makebox[2.6em][l]{$\kappa$}\parbox[t]{0.875\textwidth}{the stiffness of the phase, which comes from the
elastic stiffness of the guest chain;}
\item[$\bullet$] \makebox[2.6em][l]{$\Delta Q$}\parbox[t]{0.875\textwidth}{the misfit: the gradient $\partial_x\phi$ that the
phase would adopt if the host did not lock the registry ($g = 0$), set by the mismatch between the natural guest density of the chain inside the host, $1/c_G^{\mathrm{nat}}$ (Sec.~\ref{sec:fkmicro}), and that of the commensurate chain, $1/c_G$;}
\item[$\bullet$] \makebox[2.6em][l]{$g$}\parbox[t]{0.875\textwidth}{the locking strength, with which the host holds the chain in registry: the registry energy per unit length, $-(\kappa g/2)\cos 2\phi$, is lowest in the commensurate
registries $\phi = 0$ and $\pi$.}
\end{itemize}
For constant $\phi$, $u_j = \mathrm{const} \pm A(-1)^j$: the dimerized chain of the 4/3 approximant, with guest bonds of 2.84 and 3.03~\AA\ in GAP-Sb ($A = 0.049$~\AA). The dimerization and its amplitude are inputs; Eq.~(1) governs only their phase. The dimerization belongs to the chain: stretched to its length in the composite, the isolated chain dimerizes without the host (Fig.~1 of the Letter). The host weakens it, since the isolated chain at that length dimerizes more strongly, and its corrugation, of period $c_H/2$ for a single guest (Sec.~\ref{sec:fkmicro}), pushes successive guests, 1.5 periods apart, in opposite directions, which ties the pairing to the registry, so that a misregistry,
i.e.\ a soliton core, exchanges the pairing; the corrugation cancels for a rigid translation of the chain and does not lock it.

The phase $\phi$ is collective, not the registry of each guest with the host (the internal coordinate of superspace descriptions). In the 4/3 approximant that registry takes four values, 0, 3/4, 1/2 and 1/4 of $c_H$ from the first guest; the set is fixed by the geometry and mapped onto itself by a slip, so it belongs to the reference, together with the relaxed 4/3 approximant, the zero of Eq.~(\ref{eq:dE}) and Fig.~3(b). The phase $\phi$ adds to it a shift $\phi\,c_H/4\pi$, common to all guests between slips and larger by $c_H/4$ beyond each slip of $\pi$ (below).

A symmetry of the commensurate 4/3 structure ties $\phi$ to the registry with the host. Because $c_H = 4c_G/3$, a displacement of the guest chain by $c_G/3 = c_H/4$ equals $c_H - c_G$: it moves every guest onto the site of its neighbor, shifted by one host period, which leaves the host unchanged. The structure is therefore reproduced, with the pairing exchanged. Sliding the chain by 0.97~\AA\ thus advances $\phi$ by $\pi$, so the locking term has the period of the registry minima of Fig.~2(a), and a slip of $\pi$ removes one third of a guest, the 1/1 core of the approximants. The full displacement
from the uniform chain is therefore $u_j = (c_G/3\pi)\,\phi(x_j) + A\cos[Q_0 x_j + \phi(x_j)]$: across a slip of $\pi$ the alternation changes sign and the chain beyond it is shifted by $c_G/3$.
Fig.~2(a) shows this symmetry directly: its deepest minima are equal in energy and lie alternately 0.86 and 1.08~\AA\ apart, i.e.\ $c_G/3 \mp 2A$. At each step the chain advances by $c_G/3$ and the pairing is exchanged, so the dragged guest moves from one end of its dimer to the other; with $2A = 0.10$~\AA\ from the DFT guest bonds of that cell (2.82 and 3.02~\AA), the predicted steps are 0.87 and 1.07~\AA.

Eq.~(1) is minimized by the soliton lattice~\cite{Bak1982DevilsStaircase,BraunKivshar1998}
\begin{equation}
\phi(x) = \frac{\pi}{2} + \mathrm{am}\!\left(\frac{\sqrt{2g}}{k}\,x,\; k\right),
\label{eq:sl}
\end{equation}
in which successive slips of $\pi$ are separated by $\ell = 2kK(k)/\sqrt{2g}$ (am: Jacobi amplitude; $K$, $E$:
complete elliptic integrals; $0 < k \le 1$). Relative to the commensurate state,
\begin{equation}
\frac{\Delta E}{L} = \frac{\kappa}{\ell^2}\left[4K(k)E(k) - 2K(k)^2(1-k^2)\right] - \frac{\pi\kappa\,\Delta Q}{\ell}.
\label{eq:dE}
\end{equation}
For widely spaced solitons $\Delta E/L \approx (E_s - \pi\kappa\Delta Q)/\ell$, with $E_s = 2\kappa\sqrt{2g}$: solitons form when $\pi\kappa\Delta Q > E_s$, and their repulsion selects a finite spacing. One soliton per $n$ commensurate units gives $c_H/c_G = (4n+1)/(3n+1)$. Fig.~3(a) shows Eq.~(\ref{eq:sl}) for illustrative parameters in the strong-locking regime.

\subsection{Microscopic content of $\kappa$, $\Delta Q$ and $g$}\label{sec:fkmicro}
In the Frenkel--Kontorova picture, $\kappa$ is the stiffness of the guest chain, $\Delta Q$ is set by its natural density inside the host: the number of guests per unit length, $1/c_G^{\mathrm{nat}}$, that the dimerized chain would adopt without locking [Table~\ref{tab:fkatom}(e)], and $g$ is the corrugation of the substrate formed by the host. Table~\ref{tab:fk} (Sec.~\ref{sec:fkcalib}) compares our
atomistic estimates of these three parameters with the calibration of Fig.~3(b); Table~\ref{tab:fkatom} lists the atomistic quantities behind them, grouped by parameter, and the methods are described in Sec.~\ref{sec:fkmethods}.

The locking term of Eq.~(1) is not the potential felt by a single guest. Our GAP-Sb calculations show that a guest alone in an empty channel feels a potential of period $c_H/2 = 1.95$~\AA\ and 0.25--0.32~eV from peak to peak, but the guests of the 4/3 chain are 1.5 such periods apart, so its leading harmonic cancels along the chain. The locking is collective: sliding the 4/3 chain through the host gives an energy of period $c_G/3 = 0.97$~\AA, the period of $-g\cos 2\phi$ derived above. Its
amplitude cannot be read off the barriers of Fig.~2(a), which separate nearby metastable registries through intermediate registries and stick--slip events, i.e.\ higher harmonics that Eq.~(1) leaves out. Alternatively, we obtain it from direct sliding calculations (Sec.~\ref{sec:fkmethods}): the commensurate chain is locked by 13--21~meV per guest [Table~\ref{tab:fkatom}(h)], and the curvature of the minimum [rows (f), (g)] sets the decay of the soliton tails. The locking strength $g$ of Eq.~(1) is nevertheless treated as a calibrated parameter (Sec.~\ref{sec:fkagree}).

Since $\phi = 3\pi u/c_G$ for a chain displacement $u$, an axial stiffness $K$ (energy $\frac{1}{2}K(\partial_x u)^2$
per unit length and channel) gives $\kappa = Kc_G^2/9\pi^2$; a natural spacing $c_G^{\mathrm{nat}}$ gives $\Delta Q = 3\pi(c_G^{\mathrm{nat}} - c_G)/c_G^2$; and the curvature $V''$ of the locking potential per guest at its minimum gives
$2g = V''/(Kc_G)$, so that the tails of a soliton decay over $1/\sqrt{2g} = (Kc_G/V'')^{1/2}$.

The phonons give the stiffness of the commensurate domains, where the chain is dimerized at the spacing $c_G$ [Table~\ref{tab:fkatom}(a)--(d)]. The phase, however, varies only at the soliton cores, where the dimerization collapses and the chain is stretched, to a local spacing of 3.08--3.11~\AA\ in the relaxed approximants [row (j)]. The guest chain is strongly anharmonic. An equidistant chain, which can neither dimerize nor form slips, has the stiffness of the dimerized chain at $c_G$ (5.5 against 5.4~eV/\AA$^2$; $\kappa = 0.96$ against 0.95~eV\,\AA), but softens to 3.1 and 2.3~eV/\AA$^2$ at 3.05 and 3.09~\AA\ [row (b)], so that the phase stiffness at the cores is 0.57--0.69~eV\,\AA\ [row (d)]. The tails of a soliton, in contrast, lie in the domains: with the domain stiffness and the curvature of the locking potential they decay over 2.2 guest spacings [row (i)], close to
the 2.4--3.8 over which the registry steps in the relaxed approximants [row (k)].

\subsection{Methods for the atomistic estimates}\label{sec:fkmethods}
The quantities of Table~\ref{tab:fkatom} were obtained as follows, all with GAP-Sb.

\textit{Force constants (phonons)} [rows (a), (c), (f), (g)]. Harmonic force constants of the 4/3 approximant relaxed at 10.2~GPa (space group $P4/ncc$) were computed by finite displacements of 0.01~\AA\ (13 symmetry-distinct displacements) in $2\times2\times1$ and $2\times2\times3$ supercells (128 and 384 atoms). For wavevectors $q$ along $c$, up to $0.04 \times 2\pi/c$, we kept six collective coordinates, the uniform translations of all guests and of all host atoms along $x$, $y$ and $z$, modulated by $e^{iqz}$, and minimized the harmonic energy over all other coordinates. For the translations along $c$ this gives a $2\times2$ stiffness matrix $S(q) = S(0) + q^2 M$ per cell. Its constant part is the curvature of the energy for sliding the guests against the host, $V'' = S_{GG}(0)/8$ per guest; its $q^2$ part gives the axial stiffnesses of chain and host per channel and their
cross term, $K_G = M_{GG}/2c$, $K_H = M_{HH}/2c$ and $K_{GH} = M_{GH}/2c$ (two channels per cell of length $c$). The two supercells give $\kappa$ within 1.5\%, and the axial stiffness of chain and host together,
$K_G + K_H + 2K_{GH} = 35.3$~eV/\AA\ per channel, agrees with 35.1~eV/\AA\ from the energy of the approximant strained by $\pm1\%$ along $c$ with all atoms relaxed.

\textit{Phase stiffness} [row (d)]. The host deforms around a soliton as well, so chain and host act as springs in series, $K = (K_GK_H - K_{GH}^2)/(K_G + K_H + 2K_{GH})$, and $\kappa = Kc_G^2/9\pi^2$; $K_G$ is the spring between neighboring guests times their spacing.

\textit{Equidistant chains} [row (b)]. In a fixed host cell, the 4/3 cell of Fig.~2 at 10.2~GPa
($a = 8.036$~\AA, $c = 11.674$~\AA) repeated three, six or nine times along $c$, each channel held $N$ equally spaced guests, with $N$ chosen to give mean spacings of 2.69--3.18~\AA. The guest coordinates along $c$ were held on this uniform lattice, so that the chain can neither dimerize nor form slips; the transverse guest coordinates and all host coordinates were relaxed at fixed cell, with the center of mass of the host held along $c$. The energy was averaged over translations of the chains through the host (4 to 12 positions over $c_H/2$), and the spring $k(a) = w''(a)$ obtained from the energy per guest $w(a)$ by finite differences over neighboring spacings.

\textit{Locking} [rows (f)--(h)]. The locking of the commensurate chain was evaluated in three ways, with every coordinate other than the slide relaxed. (i) From the force constants, as the curvature $V''$ of the energy for sliding all guests uniformly against the host [row (f)]. (ii) From the force constants with the protocol of Fig.~2 (one guest held, one host atom fixed along $c$): the stiffness along the dragged coordinate is 1.56~eV/\AA$^2$ per cell; the other guests of the dragged channel follow almost rigidly (0.76--0.81 of the displacement), those of the other channel by about half, and per guest of the dragged channel, moved rigidly, the curvature is 0.47~eV/\AA$^2$ [row (g)]. (iii) Directly, by translating the relaxed 4/3 chain, or an equidistant chain at the commensurate spacing,
through the relaxed host at fixed cell, with the center of mass of each chain held: the energy is a cosine of period $c_G/3$ with an amplitude of 20.5 and 12.8~meV per guest [row (h)]. Equidistant chains at spacings away from $c_G$ (2.80--3.18~\AA) show a registry dependence of at most 0.06~meV per guest: the locking exists only where the chain is locally commensurate.

\textit{Natural spacing} [row (e)]. In the same host cells, chains of different density were translated through the host with the center of mass of each chain held and all other coordinates relaxed (the guests free to dimerize), and the energy was averaged over registry. The natural spacing $c_G^{\mathrm{nat}}$ minimizes $\Omega = E - \mu N_g$ over the guest density, with
$\mu = (E + PV)/N$ of the relaxed 4/3 approximant. $\Omega$ varies by less than 10~meV per guest between 2.92 and 3.05~\AA, so $c_G^{\mathrm{nat}}$ is uncertain by about $\pm 0.03$~\AA.

\textit{Relaxed approximants} [rows (j), (k)]. Along each guest channel of the relaxed approximants (GAP-Sb, 10.2~GPa), the displacement of each guest from the commensurate sites was split into its pair-averaged part, the registry, which steps by $c_G/3$ at each soliton, and the alternating dimerization. The registry step was fitted with a periodic $\tanh$ profile, whose decay length is the width of row (k). The local spacing of row (j) is the mean of two adjacent guest bonds; its maximum lies at the soliton core and its minimum in the domains.

\textit{Verification against DFT.} Table~\ref{tab:fkdft} compares the force-constant quantities of GAP-Sb with DFT. On the same structure and the same displacements (7~GPa, the calculations of Fig.~\ref{fig:valid}), GAP-Sb reproduces the spring of the guest chain to 10\%, but its host is 30\% stiffer and the curvature of its locking twice as large; at 10.2~GPa, each on its own relaxed structure (DFT by density-functional perturbation theory in the 32-atom cell, with the settings of Sec.~\ref{sec:dft}), the ratios are similar. The stronger locking agrees with the sliding barrier along
the DFT path of Fig.~2, 0.10~eV per cell in GAP-Sb against 0.055~eV in DFT (Sec.~\ref{sec:gap}), and with the drag stiffness for the protocol of Fig.~2, which the DFT force constants reproduce. The preferred soliton spacing is insensitive to these differences: scaling $\kappa$ and the locking of the calibrated Eq.~(1) by the DFT/GAP-Sb ratios at 10.2~GPa (0.66 and 0.47) leaves the minimum at $n = 4.4$ and changes its depth from $-0.55$ to between $-0.37$ and $-0.67$~meV/atom.

\subsection{Transverse relaxation and coupling between channels}
All our calculations relax the transverse coordinates of the guest atoms, which stay close to the channel axis: within 0.05~\AA\ (root mean square at most 0.03~\AA) along the sliding path of Fig.~2(a), against displacements of up to 2.9~\AA\ along the channel, and within 0.025~\AA\ (root mean square at most 0.02~\AA) in the relaxed approximants of Fig.~3(b), at the soliton cores as well. Transverse displacements therefore renormalize $\kappa$ and $g$ but add no second slow coordinate, and the registry phase remains a one-dimensional order parameter. Eq.~(1) describes the registry phase of one channel; the channels are coupled through the host, and all our calculations include this coupling. In the sliding landscape of Fig.~2(a), the guests of both channels of the 32-atom
cell follow the displaced atom, and the MD cells of Fig.~4 contain 640 (21/16) and 1280 (4/3) channels. The transverse coupling renormalizes $\kappa$ and $g$ and keeps the solitons of neighboring channels in register, so that the soliton lattice is three-dimensional and survives at 300~K as the atomic density wave of Fig.~4(b) (Sec.~\ref{sec:gz}).

\subsection{Calibration of Eq.~(1) on the GAP-Sb enthalpies}\label{sec:fkcalib}
Per channel, one soliton spans $N = 4n+1$ guests and $4N+1$ atoms over
$\ell = N + 1/3$ guest spacings, so that per atom
\begin{equation}
\Delta H(N) = \frac{4\ell}{4N+1}\,\tilde\kappa\left[\frac{4K(k)E(k) - 2K(k)^2(1-k^2)}{\ell^2}
            - \frac{\pi\,\Delta Q}{\ell}\right], \qquad \ell\sqrt{2g} = 2kK(k),
\label{eq:dH}
\end{equation}
with lengths in guest spacings, $\tilde\kappa$ absorbing $\kappa$ and the conversion to energy per atom, and the reference $\Delta H = 0$ the 4/3 approximant computed with the same protocol (Sec.~III B). A fit to the GAP-Sb enthalpies of $n = 1$--11, weighted by their standard errors combined with a model error of 0.009~meV/atom (chosen so that $\chi^2$ per degree of freedom is one; the scatter of the quenches is smaller than the deviation of Eq.~(\ref{eq:dH}) from the enthalpies), gives the curve of Fig.~3(b), with $\tilde\kappa = 55.8 \pm 0.5$~meV, $\Delta Q = 0.167 \pm 0.002$ and $\sqrt{2g} = 0.131$ (0.127--0.137) per guest spacing (68\% bootstrap intervals; root-mean-square deviation
0.019~meV/atom). Table~\ref{tab:fk} lists the parameters, their physical meaning and the independent
atomistic estimates of Sec.~\ref{sec:fkmicro}. The misfit has a direct physical meaning: since $\phi$ advances by $\pi$ per $c_G/3$ of displacement, without locking ($g = 0$) the chain would have a guest spacing larger than the commensurate one by $\Delta Q\,c_G/3\pi = 1.8\%$ (2.99~\AA), one slip of $\pi$ every $\pi/\Delta Q = 18.8$ guests, and a host-to-guest ratio of 1.310, close to the measured 1.3104 at 10.3~GPa~\cite{McMahon_2007}.

\begin{table}[h]
\caption{The three parameters of Eq.~(1): calibration on the GAP-Sb enthalpies [the curve of Fig.~3(b)] and independent atomistic estimates from GAP-Sb, obtained from the rows of Table~\ref{tab:fkatom} given in the last column. Cores: the stretched, undimerized chain where the phase varies.
Lengths are converted to \AA\ with $c_G = 2.935$~\AA; $\kappa = 4c_G\tilde\kappa$ (four atoms per guest spacing and channel). Calibration ranges are 68\% bootstrap intervals.}
\label{tab:fk}
\begin{ruledtabular}\small
\begin{tabular}{lp{0.24\textwidth}p{0.18\textwidth}p{0.24\textwidth}p{0.12\textwidth}}
Parameter & \raggedright Physical meaning & \raggedright Calibration & \raggedright Atomistic estimate & From \tabularnewline
\hline
$\kappa$ & \raggedright stiffness of the phase & \raggedright 0.655~eV\,\AA\ ($\tilde\kappa = 55.8 \pm 0.5$~meV) & \raggedright cores 0.57--0.69~eV\,\AA & \raggedright (d) \tabularnewline
\addlinespace
$\Delta Q$ & \raggedright misfit: phase gradient of the chain without locking & \raggedright $0.0570 \pm 0.0005$~\AA$^{-1}$ & \raggedright $0.06 \pm 0.03$~\AA$^{-1}$ & \raggedright (e) \tabularnewline
\addlinespace
$g$ & \raggedright locking strength, given as the soliton width $1/\sqrt{2g}$ (decay length of the tails) & \raggedright 7.6 (7.3--7.9) guest spacings & \raggedright 2.2 guest spacings; relaxed approximants 2.4--3.8 & \raggedright (i), (k) \tabularnewline
\end{tabular}
\end{ruledtabular}
\end{table}

\begin{table}[h]
\caption{Atomistic quantities behind Table~\ref{tab:fk}, grouped by the parameter of Eq.~(1) they determine; the methods are described in Sec.~\ref{sec:fkmethods}. All values are from GAP-Sb at 10.2~GPa. Stiffnesses are per channel, springs and curvatures per guest; $c_G = 2.935$~\AA.}
\label{tab:fkatom}
\begin{ruledtabular}\small
\begin{tabular}{lp{0.42\textwidth}p{0.27\textwidth}p{0.18\textwidth}}
 & \raggedright Quantity & \raggedright Method & GAP-Sb \tabularnewline
\hline
\multicolumn{4}{l}{\textbf{Stiffness of the phase, $\kappa$}} \tabularnewline
(a) & \raggedright spring between neighboring guests, dimerized chain at $c_G$ & \raggedright force constants (phonons) & \raggedright 5.4~eV/\AA$^2$ \tabularnewline
(b) & \raggedright spring of an equidistant chain at spacings 2.92 / 3.00 / 3.05 / 3.09~\AA & \raggedright equidistant chains & \raggedright 5.5 / 4.2 / 3.1 / 2.3~eV/\AA$^2$ \tabularnewline
(c) & \raggedright axial stiffness of the host, $K_H$ & \raggedright force constants (phonons) & \raggedright 22~eV/\AA \tabularnewline
(d) & \raggedright $\kappa$, chain and host in series: domains [(a), (c)]; cores [(b) at 3.05--3.09~\AA, (c)] & \raggedright springs in series & \raggedright 0.95; 0.57--0.69~eV\,\AA \tabularnewline
\addlinespace
\multicolumn{4}{l}{\textbf{Misfit, $\Delta Q$}} \tabularnewline
(e) & \raggedright natural guest spacing $c_G^{\mathrm{nat}}$ & \raggedright chains of different density & \raggedright $2.99 \pm 0.03$~\AA \tabularnewline
\addlinespace
\multicolumn{4}{l}{\textbf{Locking strength, $g$}} \tabularnewline
(f) & \raggedright curvature $V''$ of the locking potential at its minimum & \raggedright force constants (phonons) & \raggedright 0.68~eV/\AA$^2$ \tabularnewline
(g) & \raggedright curvature per guest for dragging one channel rigidly, protocol of Fig.~2 & \raggedright force constants (phonons) & \raggedright 0.47~eV/\AA$^2$ \tabularnewline
(h) & \raggedright amplitude of the locking for a uniform slide (cosine of period $c_G/3$) & \raggedright chains translated through the host & \raggedright 20.5 (relaxed chain), 12.8 (equidistant chain)~meV \tabularnewline
(i) & \raggedright decay length of the soliton tails, $1/\sqrt{2g} = (Kc_G/V'')^{1/2}$, domains & \raggedright from (a), (c), (f) & \raggedright 2.2 guest spacings \tabularnewline
\addlinespace
\multicolumn{4}{l}{\textbf{Solitons in the relaxed approximants}} \tabularnewline
(j) & \raggedright local guest spacing: soliton cores; domains & \raggedright relaxed approximants, $n = 3$--9 & \raggedright 3.08--3.11; 2.94--2.97~\AA \tabularnewline
(k) & \raggedright width of the registry step (decay length) & \raggedright relaxed approximants, $n = 3$--8 & \raggedright 2.4--3.8 guest spacings \tabularnewline
\end{tabular}
\end{ruledtabular}
\end{table}

\begin{table}[h]
\caption{Check of GAP-Sb against DFT for the force-constant quantities of Table~\ref{tab:fkatom} (Sec.~\ref{sec:fkmethods}). At 7~GPa both use the same DFT-relaxed structure and the same displacements (the calculations of Fig.~\ref{fig:valid}). At 10.2~GPa each uses its own relaxed structure, DFT by density-functional perturbation theory in the 32-atom cell. Stiffnesses per channel, springs and curvatures per guest.}
\label{tab:fkdft}
\begin{ruledtabular}\small
\begin{tabular}{lcccc}
 & \multicolumn{2}{c}{7~GPa, same structure} & \multicolumn{2}{c}{10.2~GPa} \\
Quantity & GAP-Sb & DFT & GAP-Sb & DFT \\
\hline
spring between neighboring guests (eV/\AA$^2$) & 5.0 & 4.6 & 5.4 & 3.9 \\
axial stiffness of the host, $K_H$ (eV/\AA) & 19.8 & 15.2 & 22.0 & 13.6 \\
$\kappa$, domains (eV\,\AA) & 0.91 & 0.74 & 0.95 & 0.63 \\
curvature $V''$ of the locking (eV/\AA$^2$) & 0.74 & 0.35 & 0.68 & 0.32 \\
drag stiffness, protocol of Fig.~2 (eV/\AA$^2$ per cell) & 1.56--1.66 & 1.06--1.18 & 1.56 & 0.98--1.13 \\
\end{tabular}
\end{ruledtabular}
\end{table}
Two of the three parameters are fixed by the position and depth of the minimum, at $N = 18.6$
(18.5--18.8), $n \approx 4.4$, and $0.553 \pm 0.005$~meV/atom. Since $\pi\Delta Q$ exceeds $2\sqrt{2g}$, an isolated soliton has a negative formation enthalpy; correspondingly, every approximant with $n \ge 3$ lies below the 4/3
reference.

\subsection{Agreement between the coarse-grained model and the atomistic calculations}\label{sec:fkagree}
Table~\ref{tab:fk} compares the calibrated parameters with the independent atomistic estimates. The misfit, which measures the natural spacing of the chain, agrees. So does the stiffness, once it is taken where the phase varies: the
gradient term of Eq.~(1) contributes only at the soliton cores, where the chain is stretched and undimerized, and the calibrated $\kappa = 0.655$~eV\,\AA\ lies within the range for the cores, 0.57--0.69~eV\,\AA, below the 0.95~eV\,\AA\ of the commensurate domains [Table~\ref{tab:fkatom}(d)].

The locking strength, in contrast, is an effective parameter. The calibrated width, 7.6 guest spacings, is two to three times the decay length of the soliton tails in the relaxed approximants, 2.4--3.8 [Table~\ref{tab:fkatom}(k)], and in the atomistic estimate, 2.2 [row (i)]. In Eq.~(1) solitons interact only through the overlap of these tails, a repulsion of short range. The enthalpies, however, contain an additional repulsion of long range: with isotropic cell relaxation the approximants keep the $c/a$ ratio of the 4/3 cell, so that the axial strain of host and chain, and its energy, grow with the density of solitons. In the approximants of Fig.~3(b) the axial stress differs from the lateral
stress by $-1.75$~GPa for $n = 1$, $-0.55$~GPa for $n = 4$ and by less than 0.1~GPa for $n \ge 9$; relaxing the cell shape lowers their enthalpy by 2.2, 0.17 and less than 0.01~meV/atom, respectively, with $c_H/a$ rising from 0.476 ($n = 1$) to 0.489 ($n = 11$). The calibration absorbs this contribution by broadening the solitons, which weakens $g$. Calibrated on the same approximants with the cell shape relaxed (Table~\ref{tab:fkcell}), the width, 3.35 guest spacings, lies within the atomistic range, while the stiffness and the natural spacing change little. The calibrated $g$ of Fig.~3(b) therefore describes the enthalpies at the cell shape of the 4/3 approximant, while the shape of a soliton is set by the atomistic stiffness and locking.

\begin{table}[h]
\caption{Calibration of Eq.~(1) on the GAP-Sb approximants with isotropic cell relaxation [Fig.~3(b)] and with the cell
shape relaxed as well (free cell), from the same starts and quench protocol (Sec.~III B), and the atomistic estimates
of Table~\ref{tab:fk}. $\kappa = 4c_G\tilde\kappa$ and $c_G^{\mathrm{nat}} = c_G(1 + \Delta Q/3\pi)$ with $c_G = 2.935$~\AA. The position
of the minimum is reliable to about $\pm 1$ in $n$: a change of 0.01~\AA\ in the natural guest spacing moves it by about
one unit.}
\label{tab:fkcell}
\begin{ruledtabular}
\begin{tabular}{lccc}
 & Isotropic cell [Fig.~3(b)] & Free cell & Atomistic estimate \\
\hline
stiffness $\kappa$ (eV\,\AA) & 0.655 & 0.633 & cores 0.57--0.69 \\
natural guest spacing $c_G^{\mathrm{nat}}$ (\AA) & 2.987 & 3.016 & $2.99 \pm 0.03$ \\
soliton width $1/\sqrt{2g}$ (guest spacings) & 7.6 & 3.35 & 2.2--3.8 \\
soliton spacing $n$ at the minimum & 4.4 & 3 & -- \\
\end{tabular}
\end{ruledtabular}
\end{table}

The position of the minimum, in contrast, cannot be obtained from atomistic parameters alone. It lies near $\pi/\Delta Q$ guests per soliton (within 4\% for both calibrations of Table~\ref{tab:fkcell}), so it follows the natural spacing of the chain, and sensitively, because the misfit is small (1.8\%): a change of that spacing by 0.01~\AA, a third of its uncertainty, changes $\Delta Q$ by a fifth and moves the minimum of Eq.~(\ref{eq:dH}) by about
one unit, from $n \approx 4.0$ to $n \approx 4.9$ at the calibrated parameters, whereas changes of $\kappa$ by 10\% or of $g$ by 20\% leave it nearly unchanged. With the atomistic estimates themselves, $\pi\Delta Q/2\sqrt{2g}$ ranges from 0.3 to 1.6 within their uncertainties, across the commensurate--incommensurate threshold at 1, so they cannot decide even whether solitons form; the enthalpies decide it. Correspondingly, the minimum lies at $n \approx 4.4$ with the cell shape of the 4/3 approximant and at $n \approx 3$ with the cell shape relaxed (Table~\ref{tab:fkcell}), near the experimental value $n \approx 4.5$ (300~K), which follows from the measured host-to-guest ratio 1.3104~\cite{McMahon_2007}. Locating it more precisely would require an accurate description of the elastic response of the host and guest and of the misfit between them. GAP-Sb, which contains no configuration of
approximant size in its training set, recovers the instability of the commensurate structure against solitons and their preferred intermediate density; we do not regard the exact position of the minimum as a prediction.

Narrow solitons with a repulsion $\propto 1/\ell$, and the $g \to 0$ limit, a sinusoidal modulation, fit somewhat less well (0.036 against 0.019~meV/atom). The enthalpies thus fix the formation enthalpy and the repulsion of the solitons, and with them the spacing, but they do not by themselves distinguish a soliton lattice from a sinusoidal modulation; that distinction comes from the narrow cores of Fig.~3(c--e) and from the parity exchange of Sec.~\ref{sec:gz}.

\clearpage
\section{Diffraction patterns}\label{sec:xrd}
One-dimensional x-ray powder patterns were computed directly from the atomic configurations with an in-house code,without assuming a superspace model. The kinematic intensity per atom, $f(q)^2|\sum_j \exp(i\mathbf{q}\cdot
\mathbf{r}_j)|^2/N$, was evaluated at the reciprocal-lattice vectors of the MD supercell with $|\mathbf{q}| \le 5$~\AA$^{-1}$ and averaged over orientations. 

The patterns were broadened with a pseudo-Voigt function of full width at half maximum 0.015~\AA$^{-1}$ and Lorentzian fraction 0.5, matched to the isolated peaks of the measured profile; the broadening conserves the integrated intensity. The atomic form factor $f(q)$ of Sb is that of the Cromer--Mann parameterization. Because the supercells are long along $c$ (1876~\AA\ for the 21/16 structure, 646~\AA\ for the 4/3 reference; Sec.~\ref{sec:md}), their reciprocal lattices sample the main, satellite and second-order reflections without any indexing. The patterns were computed from the last 300~K configuration of each run (Sec.~\ref{sec:md}) and are compared with the one-dimensional profile of Sb-II at 10.2~GPa provided by M.~I.~McMahon, primarily in peak positions. No Lorentz--polarization or absorption correction is applied, and each pattern is normalized to its strongest reflection, $(2\,1\,1\,0)$ at 2.37~\AA$^{-1}$.

Before the patterns are computed, each simulated configuration is scaled affinely to the experimental lattice, in the
plane by $a_\mathrm{exp}/a_\mathrm{sim}$ and along the chains by $c_{H,\mathrm{exp}}/c_{H,\mathrm{sim}}$ ($a = 8.0676$~\AA,
$c_H = 3.9071$~\AA~\cite{McMahon_2007}): by 1.0047 and 1.0001 for the 21/16 structure and by 1.0063 and 0.9925 for the
4/3 approximant. This places the main reflections at their measured positions without any adjustable parameter; the host-to-guest ratio, and with it the positions of the satellite reflections, are those of the simulation. A smooth background was removed from the measured profile and from the simulated patterns with the same peak-clipping algorithm (SNIP~\cite{sm-Ryan1988}, window $\pm 0.06$~\AA$^{-1}$); in the simulations it consists of the thermal diffuse scattering, in the measurement it also contains the scattering from the pressure cell. 
A measured guest or modulation reflection counts as reproduced if a simulated peak of at least a quarter of its measured height lies within 0.01~\AA$^{-1}$ of it: the 21/16 structure reproduces all eight labeled in Fig.~4(c--f) (mean offset 0.003~\AA$^{-1}$, largest 0.009), the 4/3 approximant none.
The profile was measured on ID09 at the ESRF with a wavelength of 0.4157~\AA~\cite{McMahon_2007}; the simulated patterns are plotted for the same wavelength.

\bibliography{SbII}